\documentstyle[here,epsfig]{mn2e}
\def\simlt{\mathrel{\rlap{\lower 3pt\hbox{$\sim$}}\raise 2.0pt\hbox{$<$}}}
\def\simgt{\mathrel{\rlap{\lower 3pt\hbox{$\sim$}} \raise 2.0pt\hbox{$>$}}}

\def\Msun{M_{\odot}}

\def\gtsima{$\; \buildrel > \over \sim \;$}\def\gtsima{$\; \buildrel > \over
  \sim \;$}

\def\ltsima{$\; \buildrel < \over \sim \;$}
\def\gtrsim{\lower.5ex\hbox{\gtsima}}
\def\lesssim{\lower.5ex\hbox{\ltsima}}

\newcommand{\q}{\begin{equation}}
\newcommand{\qa}{\begin{eqnarray}}
\newcommand{\qs}{\begin{eqnarray*}}
\newcommand{\nq}{\end{equation}}
\newcommand{\nqa}{\end{eqnarray}}
\newcommand{\nqs}{\end{eqnarray*}}

\begin{document}

\title[BSSs in dSphs] 
{Blue straggler stars in dwarf spheroidal galaxies}

\author[M. Mapelli et al.]{M. Mapelli$^{1}$, E. Ripamonti$^{2}$, E. Tolstoy$^{2}$, S. Sigurdsson$^{3}$, M. J. Irwin$^{4}$, G. Battaglia$^{2}$
\\
$^1$Institute for Theoretical Physics, University of Z\"urich, Winterthurerstrasse 190, CH--8057 Z\"urich, Switzerland; {\tt
mapelli@physik.unizh.ch}\\ $^2$Kapteyn Astronomical Institute, University of Groningen, Postbus 800, 9700 AV Groningen, the Netherlands\\ $^3$Department of Astronomy and Astrophysics, The
Pennsylvania State University, 525 Davey Lab, University Park,
PA~16802, US\\$^{4}$Royal Greenwich Observatory, Madingley Road, Cambridge CB3 0EZ, UK\\}

\maketitle 
\vspace {7cm}

\begin{abstract}

Blue straggler star (BSS) candidates have been observed in all old dwarf
spheroidal galaxies (dSphs), however whether or not they are authentic
BSSs or young stars has been a point of debate. To both address this issue and
obtain a better understanding of the formation of BSSs in different
environments we have analysed a sample of BSS candidates in two nearby
Galactic dSphs, Draco and Ursa Minor. We have determined their radial
and luminosity distributions from wide field multicolour imaging data
extending beyond the tidal radii of both galaxies.  

BSS candidates are uniformly distributed through the host galaxy,
whereas a young population is expected to show a more clumpy
distribution.
Furthermore, the observed radial distribution of BSSs,
normalized to both red giant branch (RGB) and horizontal branch (HB)
stars, is almost flat, with a slight decrease towards the centre.
Such a distribution is at odds with the predictions for a young stellar
population, which should be more concentrated.  Instead, it is
consistent with model predictions for BSS formation by mass transfer in
binaries (MT-BSSs).  Such
%The observed spatial and luminosity distributions of BSSs are consistent with model predictions for formation by mass transfer in binaries (MT-BSSs), but not for formation models assuming stellar collisions (COL-BSSs).  Our
results, although not decisive, suggest that these candidates are indeed BSSs and that MT-BSSs form
in the same way in Draco and Ursa Minor as in globular clusters. This
favours the conclusion that Draco and Ursa Minor are truly `fossil'
galaxies, where star formation ceased completely more than 8 billion
years ago.
  
\end{abstract}

\begin{keywords}
blue stragglers - stellar dynamics - galaxies: dwarf - galaxies: individual: Draco - galaxies: individual: Ursa Minor
\end{keywords}

\section{Introduction}

Blue straggler stars (BSSs) are stars located above and blue-ward of the
 main sequence (MS) turn-off 
in a color-magnitude diagram (CMD). They were first
discovered in a globular cluster (M3, Sandage 1953), and are mainly
observed in star clusters (Fusi Pecci et al. 1992; Ferraro et al. 1993, 1997; Zaggia, Piotto \& Capaccioli 1997; Ferraro et al. 2003, 2004; Sabbi et al. 2004; Hurley et al. 2005; Lanzoni et al. 2007a, 2007b and references therein), where the tiny (if any)  spread
in the stellar age makes their identification straight forward. 
However, there have been attempts to find halo BSSs in the Milky Way,
which show up as high-velocity stars brighter and hotter than turnoff
stars in the Galactic halo (Carney et al. 2001; Carney, Latham \& Laird
2005; Beers et al. 2007).

Dwarf spheroidal galaxies (dSphs) seem
natural places to search for BSSs. Mateo et al. (1991) and Mateo,
Fischer \& Krzeminski (1995) first indicated the existence of a large
number of stars brighter than the turn-off mass in the Sextans
dSph. Mateo et al. (1995) suggested that these stars might be ordinary
MS stars substantially younger than the bulk of
the other stars. 
%THIS SENTENCE MAKES YOU WONDER WHY YOU BOTHERED WITH THIS STUDY?
%However, recent studies tend to discard this interpretation
%identifying most of these stars with BSS (Lee et al. 2003
%and references therein; hereafter L03).  
BSS candidates have been found in varying numbers in most dSphs, such as
Sculptor (e.g., Hurley-Keller, Mateo \& Grebel 1999; Monkiewicz et
al. 1999), Draco (Aparicio, Carrera \& Mart\'inez-Delgado 2001;
hereafter A01) and Ursa Minor (Carrera et al. 2002; hereafter C02).

The issue of whether these stars are young or BSSs has not been
quantitatively addressed because there were not suitable models of
BSS formation with predictive power with which to compare the observations.
This also means that our proper understanding of
the star formation history of these `predominantly old' dSphs remains
in doubt. Has there actually 
been low-level star formation in the last
8$-$10 Gyr in these galaxies or can they really be considered `fossil'
galaxies? 
Thus, in order to reconstruct the star formation history of
dSphs, it is crucial to understand whether the observed BSS candidates
are true BSSs rather than younger stars.

A second unsolved question about BSSs is their formation mechanism
itself. BSSs are believed to have been somehow refuelled with hydrogen
after the MS phase. However, the refuelling mechanism is still
unknown. It has been proposed (McCrea 1964) that mass transfer in
isolated binaries can lead to the formation of BSSs. On the other hand,
BSSs could be the products of stellar collisions, occurring during (or
triggered by) three- and four-body encounters (Davies, Benz \& Hills
1994; Sigurdsson, Davies \& Bolte 1994; Lombardi et al. 2002). In the
first case, i.e. mass-transfer BSSs (MT-BSSs), BSSs can form if
binaries are allowed to quietly evolve until they start the mass-transfer 
phase. This implies that the local density should not be too
high, otherwise gravitational interactions will perturb the mass
transfer.  In the second case, i.e. collisionally born BSSs (COL-BSSs),
the density must be sufficiently high to guarantee a short collision
time-scale.

In globular clusters both processes might occur, as in the
high-density core we find the perfect conditions for COL-BSS formation,
while in the periphery MT-BSSs can originate from isolated binary
evolution. In some circumstances, the features of the observed BSS
population can be explained only by invoking a joint contribution by
these mechanisms (Leonard 1989; Fusi Pecci et al. 1992;
Bailyn \& Pinsonneault 1995; Ferraro et al. 1997; Sills \& Bailyn 1999;
Hurley et al. 2001). For example, in some globular clusters the
BSS radial distribution is bimodal (Ferraro et al. 1993, 1997;
Zaggia et al. 1997; Ferraro et al. 2004; Sabbi et al. 2004), with a central peak, a minimum at intermediate radii, and a further rise at the
periphery. Dynamical simulations by Mapelli et al. (2004, 2006, hereafter M04, M06, respectively) showed that this bimodal distribution can be
reproduced only by requiring  the central BSSs to be
mainly COL-BSSs,  and the peripheral BSSs  to be
MT-BSSs. Also the luminosity function of BSSs in some globular clusters
(Bailyn \& Pinsonneault 1995; Sills \& Bailyn 1999; Sills et al. 2000;
Ferraro et al. 2003; Monkman et al. 2006) suggests the coexistence of
COL-BSSs and MT-BSSs.
%?? (generally fainter than the previous ones).

In dSphs the collisional formation of BSSs should be impossible, as
the central density of these galaxies never reaches sufficiently high
values to allow stellar encounters. Thus, we expect BSSs in dSphs to be
solely MT-BSS type.

In this paper we seek to quantify the potential BSS population characteristics 
in dSphs and thereby learn more about BSS formation and evolution, and 
also about the star formation history in dSphs.
First of all, we  check whether the main
characteristics (such as radial and luminosity distribution) of BSS
candidates in dSphs are more consistent with those predicted by theoretical
models of BSSs, or if they are more similar to MS
young stars.  At the same time, we would like to test whether the
proposed formation mechanisms for BSSs also work in dSphs,
i.e.  whether the BSS candidates in dwarf galaxies can be
connected with MT-BSSs.

We focus on the BSS population of two dSphs, Ursa Minor and Draco (see Table~1).
These are among the faintest and most diffuse dwarfs in the Local Group
(Irwin \& Hatzidimitriou 1995, hereafter IH95; Mateo 1998). They also appear to be among the most dark matter dominated objects we
know about (Kleyna et al. 2003; Wilkinson et al. 2004).  Ursa Minor and
Draco are also among those dSphs of the Local Group where star formation
%has been the least efficient and - not really !
appeared to cease early on ($\gtrsim{}8-10$ Gyr
ago, see Mateo 1998; Hernandez, Gilmore \& Valls-Gabaud 2000; A01; C02;
Bellazzini et al. 2002). In both these galaxies a significant number of
BSS candidates have been detected (A01; C02). If the observed BSS 
candidates in these two dSphs can be explained as young MS stars,
the existence of BSSs in dwarf galaxies can probably be safely rejected. 
However, if instead these stars do behave like authentic BSSs, then Ursa Minor 
and Draco should be really considered two 'fossil' galaxies, where star 
formation indeed completely stopped many Gyr ago.  Furthermore, by studying
such diffuse systems we can also learn about the
properties of BSSs in a much less dense environment than a globular
cluster.

%Given the poor star formation history of Ursa Minor and Draco, if the BSS candidates in Ursa Minor and Draco were shown to be cannot be e

In Section 2 we present the data on which our analysis is based
. In Section 3 we discuss the observational features of
BSS candidates, with particular emphasis on the luminosity function and radial
distribution characteristics. In Section 4 we
describe our dynamical simulations and compare the results with
observations. A comparison with previous work on BSS
candidates in dwarf galaxies and globular clusters, is presented in
Section 5. Finally in Section 6 we present our conclusions.

\section{The data}
%%%%%%%%%%%%%%%%%%%%%%%%%%%%%%%%%%% FIGURE 1 %%%%%%%%%%%%%%%%%%%%%%%%%%%%%%%%%%
\begin{figure*}
\center{{
\epsfig{figure=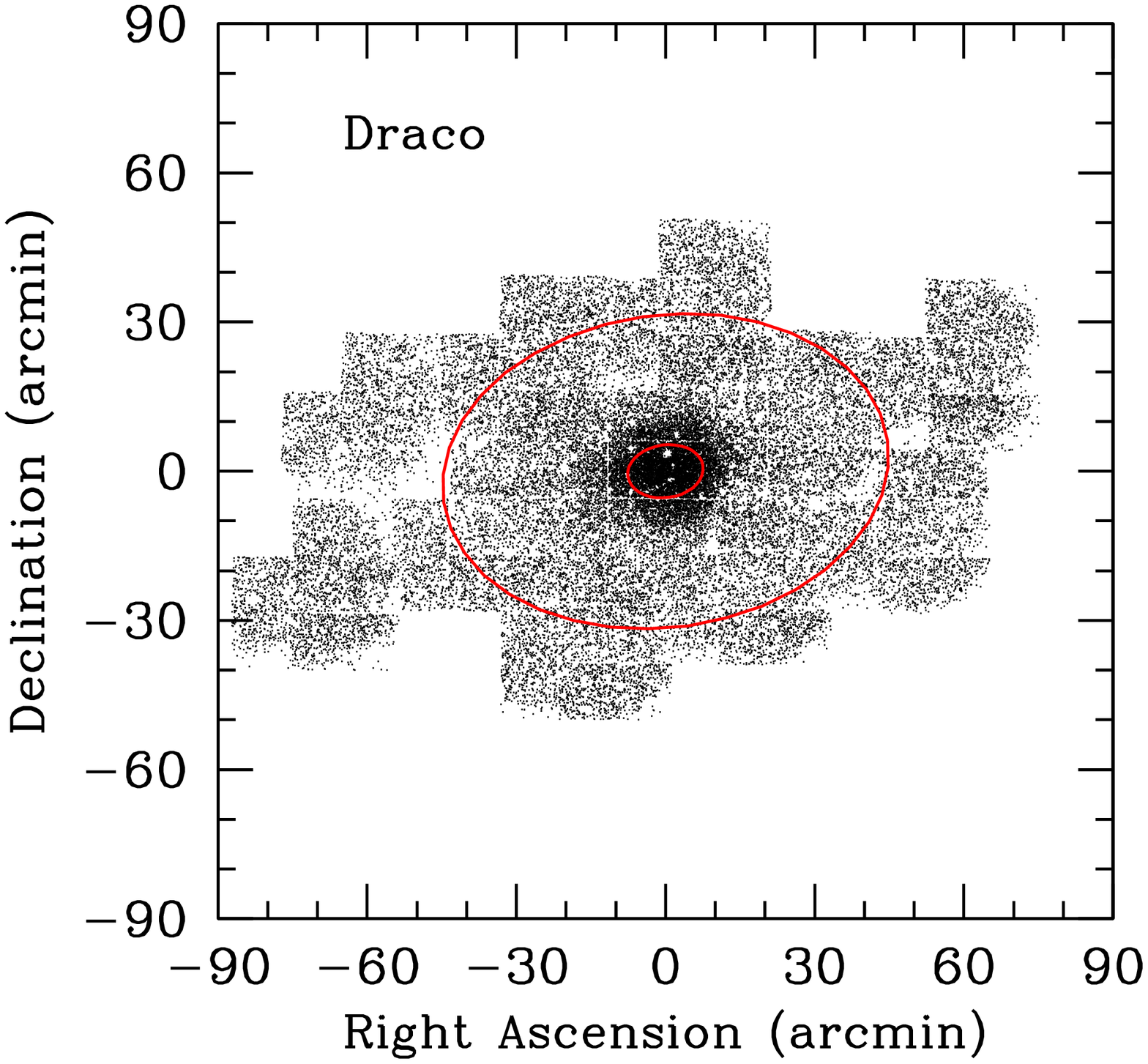,height=8cm}
\epsfig{figure=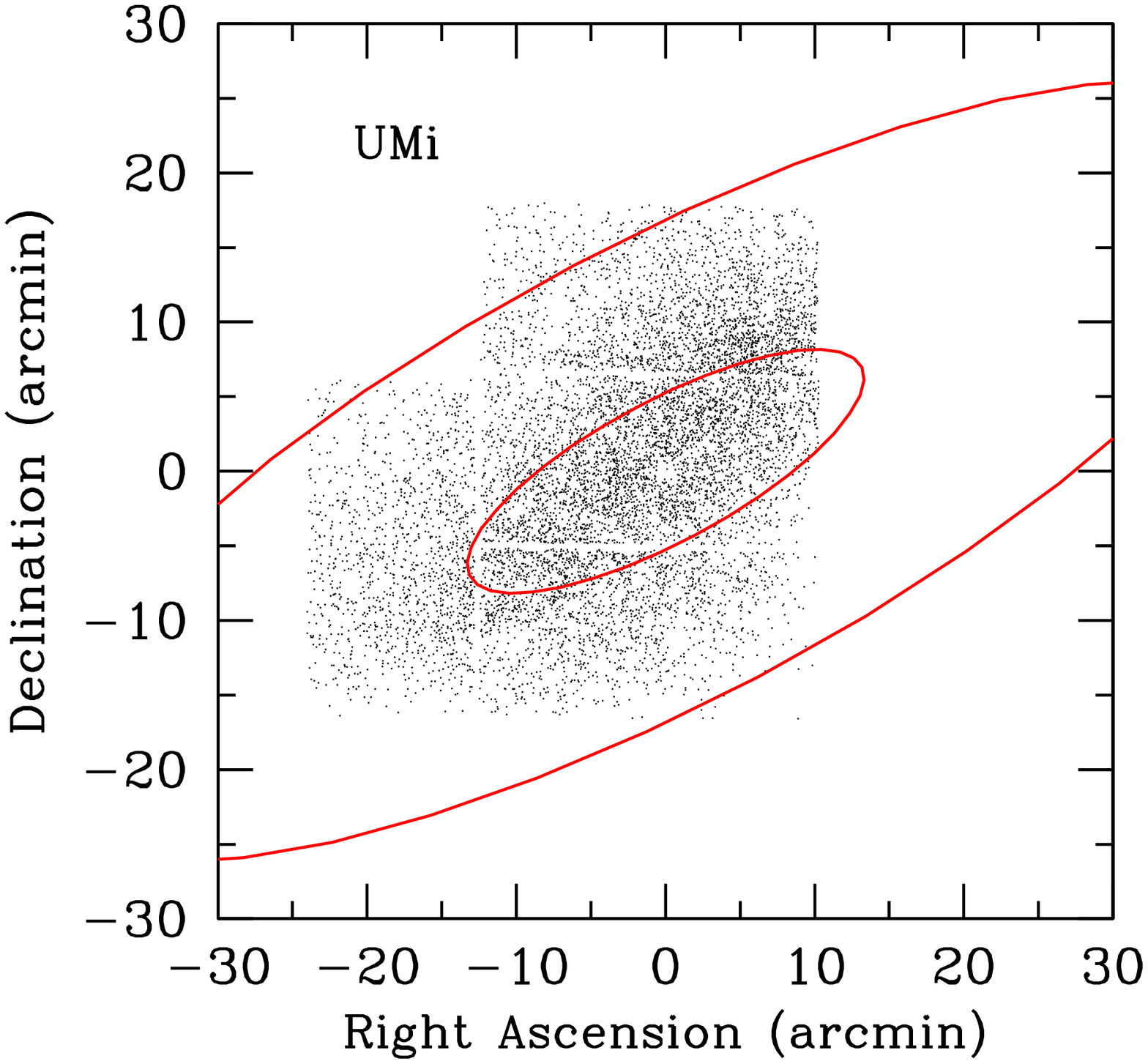,height=8cm}
\epsfig{figure=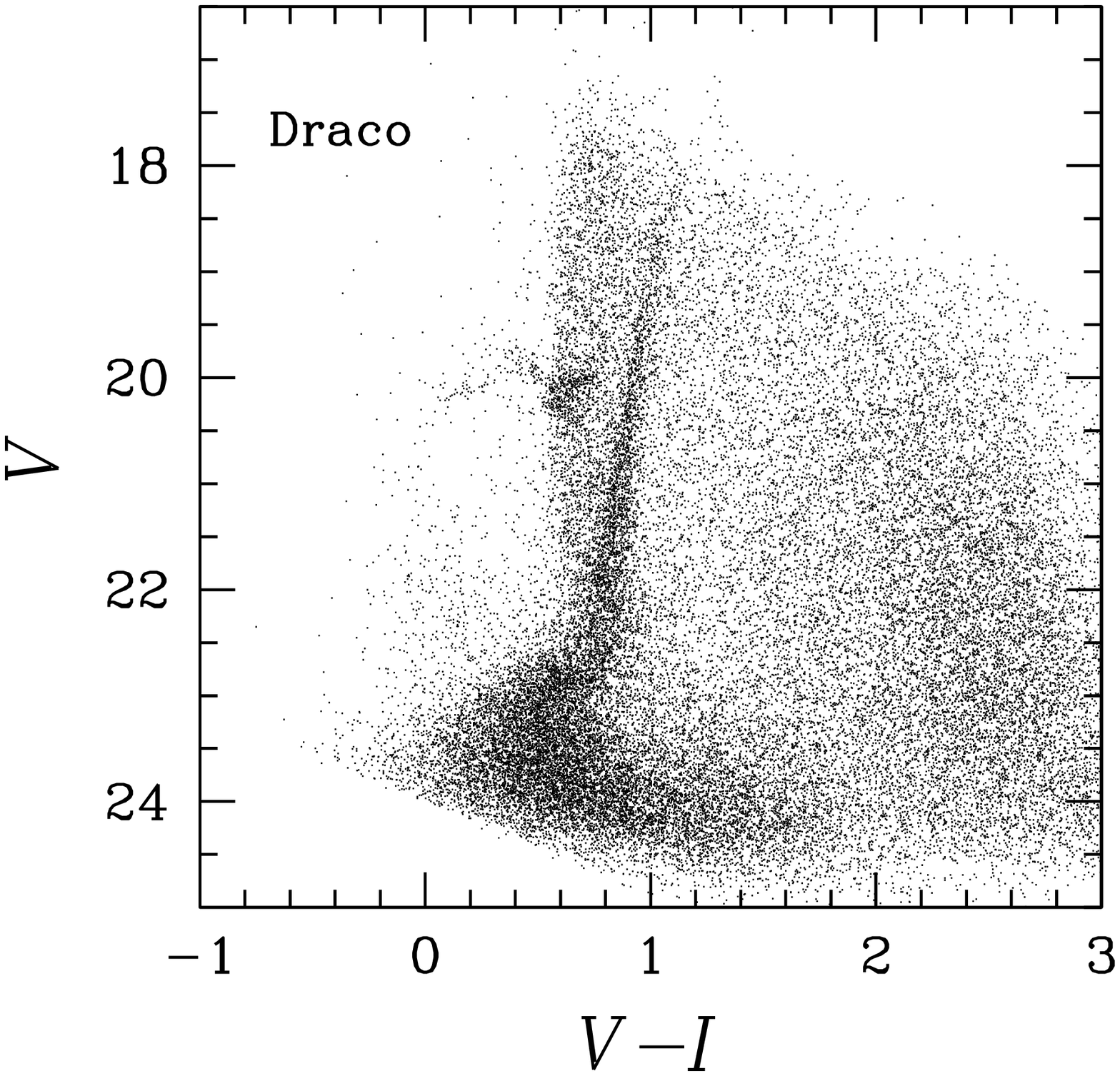,height=8cm}
\epsfig{figure=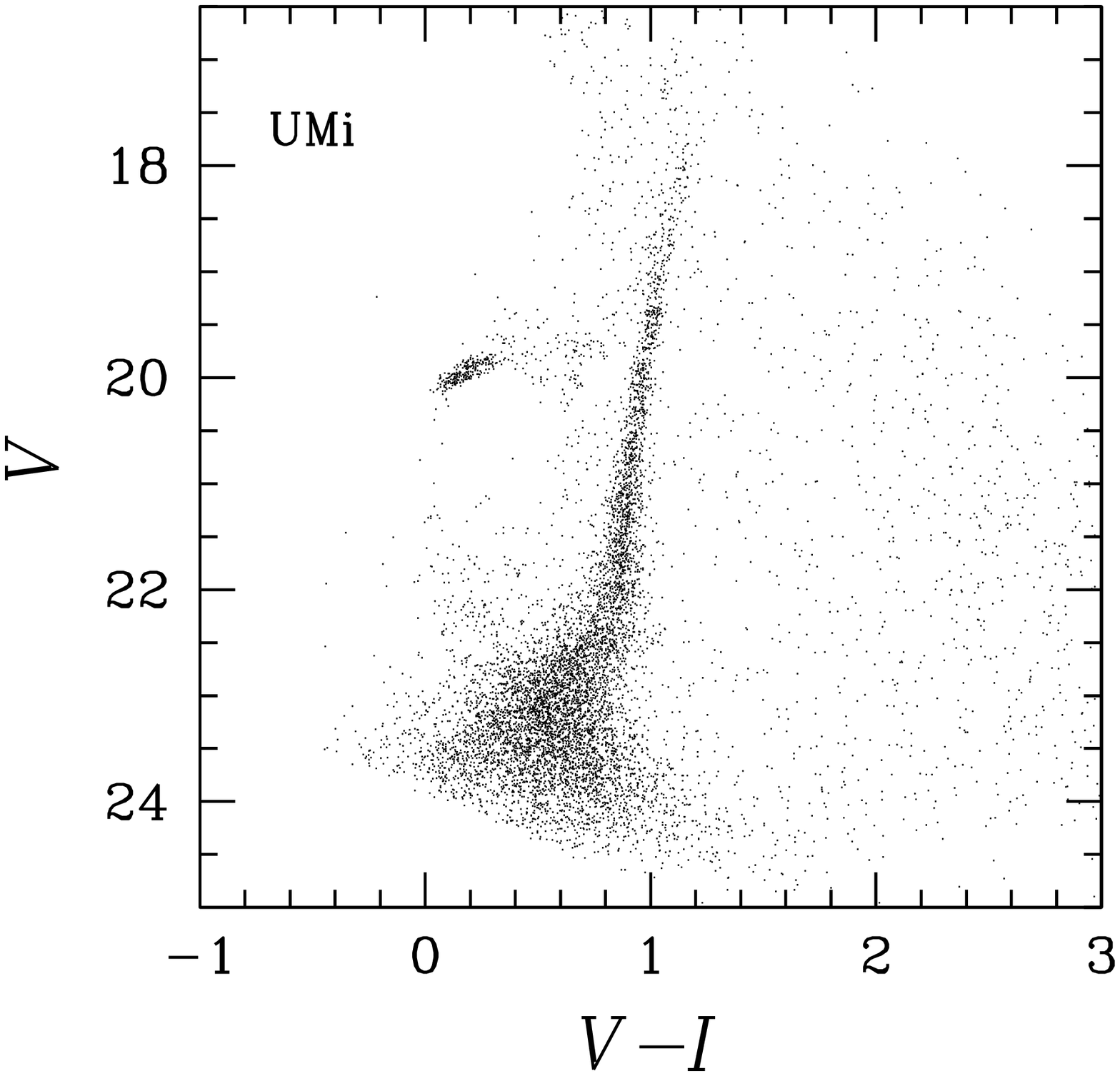,height=8cm}
}}
\caption{\label{fig:spatial} Upper panel: Right Ascension and
Declination of the stars imaged in Draco (left) and Ursa Minor
%these are presumably \deltaRA *cos(dec) and \deltaDec
(right). The concentric ellipses indicate tidal and core radii ($r_t$
% you can't see the core ellipse for Draco
and $r_c$; the adopted values are listed in Table 1).
%; for Draco $r_c=7.63$ arcmin, $r_t=45.1$ arcmin (S\'egall et al. 2007) and for  Ursa Minor $r_c=15.8$ arcmin, $r_t=50.6$ arcmin (IH95). 
In both cases the origin of the axes coincides with the centre
of the observed galaxy. Bottom panel: CMDs of Draco (left) and Ursa Minor
(right).}
\end{figure*}
%%%%%%%%%%%%%%%%%%%%%%%%%%%%%%%%%%%%%%%%%%%%%%%%%%%%%%%%%%%%%%%%%%%%%%%%%%%%%%%
\subsection{INT/WFC survey data}
%I.e., where do they come from, published/unpublished, errors etc. etc.
The Isaac Newton Telescope (INT) Wide Field Camera (WFC) is a mosaic of four 4k $\times$ 2k EEV chips, 
offering a field of view of $\sim$0.29 square degrees. It is mounted in the 
prime focus of the 2.5-m Isaac Newton Telescope on La Palma.  The $V$'-band 
(Harris filter) and $i$'-band (SDSS-like)\footnote{For filter responses see
{\tt http://www.ast.cam.ac.uk/$\sim{}$wfcsur/technical/filters/}} data extend 
beyond the tidal radius in both Draco and Ursa Minor
(see the upper panels of Fig.~\ref{fig:spatial}). They were taken
as part of the INT Wide Field Survey (McMahon et al 2001) during 2002 with
an average seeing of 1~arcsec and in generally photometric conditions.
The images were processed in the standard way with the INT WFC pipeline
(Irwin \& Lewis 2001).  The two-dimensional instrumental signature
removal includes provision for: non-linearity correction at the detector
level; bias and overscan correction prior to trimming to the active
detector areas; flatfielding; and fringe removal in the $i$'-band.

Catalogue generation follows the precepts outlined by Irwin (1985, 1996)
and includes the facility to: automatically track any background variations
on scales of typically ~20-30 arcsec; detect and deblend images or groups
of images; and parameterise the detected images to give various (soft-edged)
aperture fluxes, position and shape measures.  The generated catalogues 
start with an approximate World Coordinate System (WCS) defined by the 
known telescope and camera properties (e.g. WCS distortion model) and are 
then progressively refined using all-sky astrometric catalogues [e.g.  United States Naval Observatory (USNO) catalogue of astrometric standards, 
Automated Plate Measuring Machine (APM) catalogue, Two Micron All Sky Survey (2MASS)] to give internal precision generally better than 0.1 arcsec and 
global external precision of ~0.25 arcsec or better.  These latter numbers 
are solely dependent on the accuracy of the astrometric catalogues used in 
the refinement.

All catalogues for all CCDs for each pointing are then processed using
the image shape parameters for morphological classification in the main
categories: stellar; non-stellar; noise-like.  A sampled curve-of-growth
for each detected object is derived from a series of aperture flux
measures as a function of radius.  The classification is then based on
comparing the curve-of-growth of the flux for each detected object with
the well-defined curve-of-growth for the general stellar locus.  This
latter is a direct measure of the integral of the point spread function
(PSF) out to various radii and is independent of magnitude, if the
data are properly linearised, and if saturated images are excluded.  The
average stellar locus on each detector is clearly defined and is used as
the basis for a null hypothesis stellar test for use in classification.
The curve-of-growth for stellar images is also used to automatically
estimate frame-based aperture corrections for conversion to total flux.

The photometric standards observed during the run (mainly Landolt 1992
and spectrophotometric standards) are automatically located in a standards
database and used to estimate the zero-point in each passband for every
pointing containing any of these standards.  The trend in the derived
zero-points is then used to assign a photometric quality index for
each night and also as a first pass estimate for the magnitude calibration
for all the observations.  

Various quality control plots are generated by the pipeline and these
are used to monitor characteristics such as the seeing, the average
stellar image ellipticity (to measure trailing), the sky brightness
and sky noise, the size of aperture correction for use with the 'optimal'
aperture flux estimates (here 'optimal' refers to the well-known property
that soft-edged apertures of roughly the average seeing radius provide
close to profile fit accuracy, e.g. Naylor 1998). 

The `optimal' catalogue fluxes for the $V$', $i$' filters for each 
field are then combined to produce a single matched catalogue 
for each pointing and the overlaps between pointings are used to 
cross-calibrate all the catalogues to a common system with typical accuracy
1-2 per cent across the survey region.  The final step is to produce a unique 
catalogue for the whole region by removing spatially coincident (within 1 
arcsec) duplicates.

As a final stage the data are converted\footnote{For details of the conversion 
see {\tt http://www.ast.cam.ac.uk/$\sim{}$wfcsur/technical/photom/}} from the instrumental 
WFC $V$' and $i$' passbands to the Johnson-Cousins $V$,$I$ system to obtain the 
standard CMD shown in the lower panel of
Fig.~\ref{fig:spatial}.

%%%%%%%%%%%%%%%%%%%%%%%%%%%%%%%%%%% FIGURE 2 %%%%%%%%%%%%%%%%%%%%%%%%%%%%%%%%%%
\begin{figure*}
\center{{
\epsfig{figure=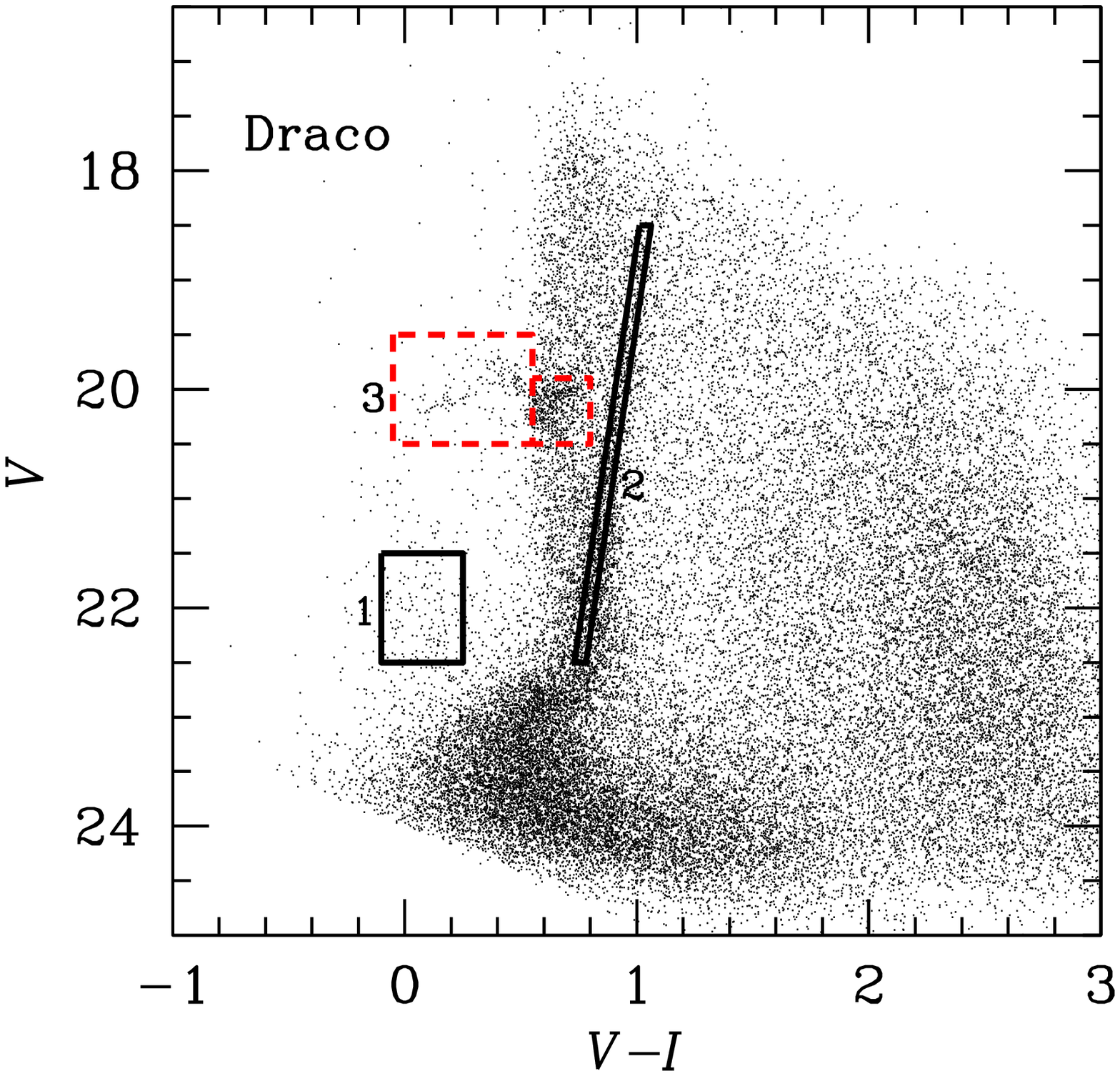,height=8cm}
\epsfig{figure=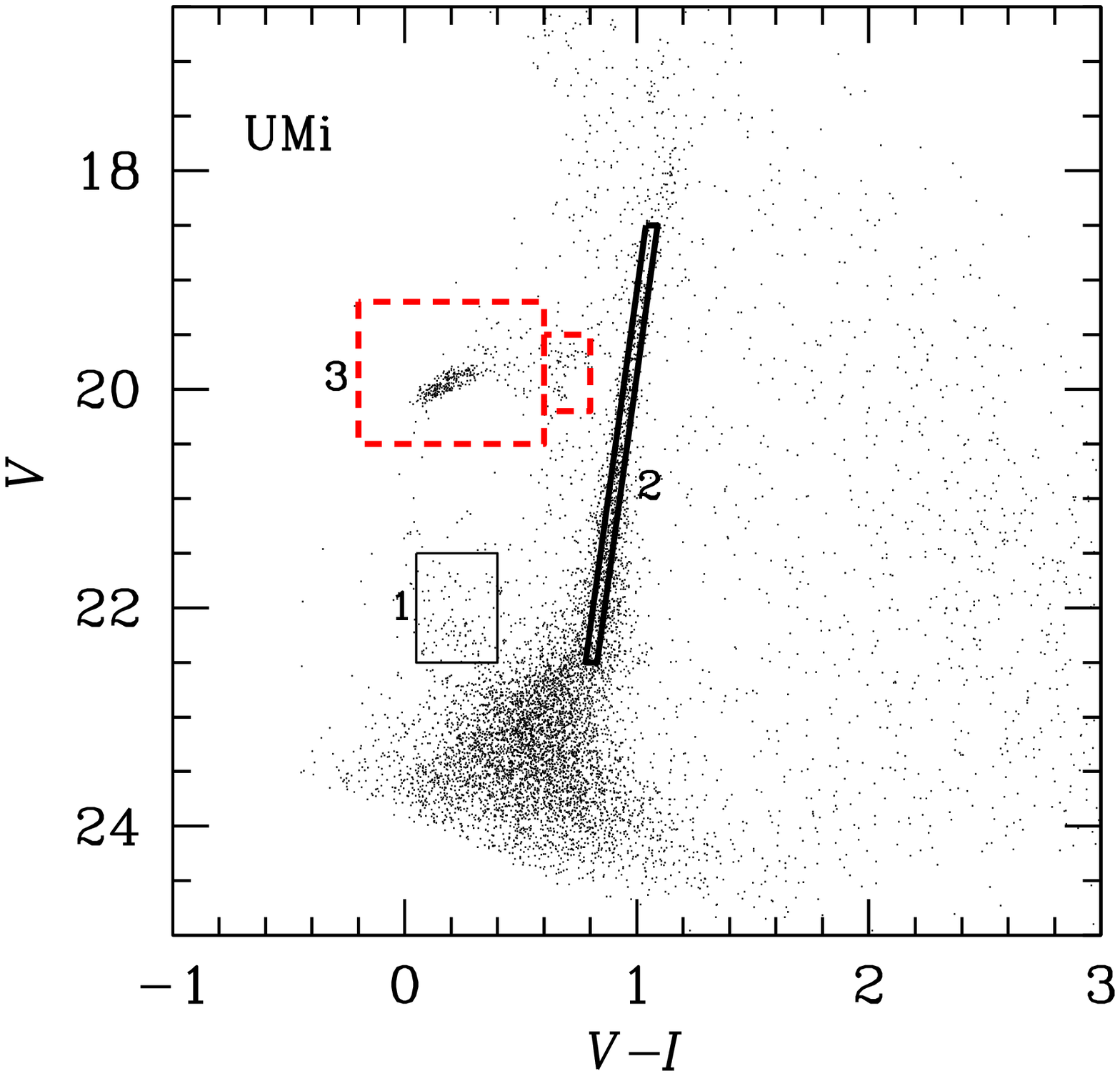,height=8cm}
}}
\caption{\label{fig:fig1} CMD of Draco (left panel) and Ursa Minor
(right) with stellar population selection boxes overlaid. Boxes indicated 
by the solid line and labelled as 1 and 2 are
the BSSs and RGB stars, respectively. Boxes indicated by the dashed line
and labelled as 3 are the HB stars, divided as blue and red.}
\end{figure*}
%%%%%%%%%%%%%%%%%%%%%%%%%%%%%%%%%%%%%%%%%%%%%%%%%%%%%%%%%%%%%%%%%%%%%%%%%%%%%%%

\subsection{Stellar population selection criteria}

From these data 
%in addition to samples of {\bf contaminating} foreground
%stars\footnote{{\bf Our data are also contaminated both by Milky Way
%stars in the foreground, and by extragalactic objects (e.g. quasars and
%unresolved galaxies) in the background. Since the foreground component
%is dominant, in the rest of this paper we will refer to any type of
%contamination as ``foreground'', unless the distinction is important.}}
we selected three different populations: BSSs, red giant branch (RGB)
and horizontal branch (HB) stars. The last two populations are
considered good tracers of the overall light from the galaxy, and we use them as a
comparison for BSS distributions. For all these populations we adopt
more conservative selection criteria with respect to previous works
(see e.g. A01; C02; Lee et al. 2003, hereafter L03), in order to
minimize the contamination by stars of different populations. The
regions of the CMD we associate with BSSs, RGB and HB stars are
indicated in Fig.~\ref{fig:fig1} as boxes 1, 2 and 3, respectively.  In
particular, for BSSs we chose the $V$ and ($V$-$I$) range to be (i)
sufficiently above the turn-off, in order to avoid contamination from
the MS, (ii) blue-ward of the RGB, avoiding not only contamination from
these stars but especially the region of the CMD most affected by the
Galactic foreground, (iii) red-ward of a possible faint extension of the
very blue extended HB.
 
Furthermore, we select a narrow strip of RGB stars (box 2), to limit the
effect of binaries and errors in magnitude. The large number of RGB
stars in the sample allows us this conservative choice. Finally, the HB
region is divided in two different regions, approximately corresponding
to the red HB (RHB) and the blue HB (BHB). As it has already been noted
(C02), the HB in Ursa Minor is substantially bluer than in Draco.

We also checked that these more restrictive selection criteria do not
significantly affect our results both for the radial and for the
luminosity distribution (see next section for the comparison with L03).

%%%%%%%%%%%%%%%%%%%%%%%%%%%%%%% TABLE 1%%%%%%%%%%%%%%%%%%%%%%%%%%%%%%%%%
\begin{table*}
\begin{center}
\caption{Galaxy parameters} \leavevmode
\begin{tabular}[!h]{lllllllll}
\hline
Galaxy
& $d$$^{\rm a}$ (kpc)
& $r_{c}$$^{\rm b}$ (arcsec)
& $r_{t}$$^{\rm b}$ (arcsec)
& $\sigma_c$ (km s$^{-1}$)$^{\rm c}$
& $n_{c}$ (stars pc$^{-3}$)$^{\rm d}$
& $W_0$$^{\rm e}$
& $c$$^{\rm e}$
& ellipticity$^{\rm b}$ \\
\hline
Draco      & 83 & 457.8 & 2706 & 10.5 & $3.2\times{}10^{-3}$ & 2.0 &  0.76  & 0.29 \\
Ursa Minor & 76 & 948 & 3036 & 12.5 & $10^{-3}$ & 0.45 &  0.52  & 0.56 \\
\noalign{\vspace{0.1cm}}
\hline
\end{tabular}
\end{center}
\footnotesize{ $^{\rm a}$ We assume distance moduli of 19.60 (Draco) and
  19.41 (Ursa Minor); see appendix B; $^{\rm
b}$Core radius ($r_c$), tidal radius ($r_t$) and ellipticity are from
S\'egall et al. (2007) and from IH95 for Draco and Ursa Minor,
respectively.  $^{\rm c}$Core velocity dispersion of the dSph, from
Wilkinson et al. (2004).  $^{\rm d}$Core density (n$_c$)of the dSph,
derived from our data.  $^{\rm e}$Central adimensional potential ($W_0$)
and concentration [$c={\rm log}(r_c/r_t)$] are derived from our simulations. $c$ is
consistent with IH95 for Ursa Minor and with
S\'egall et al. (2007) for Draco.  }
\end{table*}
%%%%%%%%%%%%%%%%%%%%%%%%%%%%%%%%%%%%%%%%%%%%%%%%%%%%%%%%%%%%%%%%%%%%%%%%%%%%%
%%%%%%%%%%%%%%%%%%%%%%%%%%%%%%%%%%% FIGURE 2 %%%%%%%%%%%%%%%%%%%%%%%%%%%%%%%%%%
%\begin{figure*}
%\center{{
%%\epsfig{figure=draco_surfdensity.ps,height=8cm}
%%\epsfig{figure=ursaminor_surfdensity.ps,height=8cm}
%\epsfig{figure=radial_distr_draco.eps,height=8cm}
%\epsfig{figure=ursaminor_surfdensity.ps,height=8cm}
%}}
%\caption{\label{fig:fig2} Surface density of BSS (dotted line connecting circles), HB
%(dashed line connecting squares) and RGB stars (solid line  connecting triangles) in Draco (left panel) and Ursa Minor
%(right). Size of points and poissonian errors are comparable.
%}
%\end{figure*}
%%%%%%%%%%%%%%%%%%%%%%%%%%%%%%%%%%%%%%%%%%%%%%%%%%%%%%%%%%%%%%%%%%%%%%%%%%%%%%%
%%%%%%%%%%%%%%%%%%%%%%%%%%%%%%%%%%% FIGURE 3 %%%%%%%%%%%%%%%%%%%%%%%%%%%%%%%%%%
\begin{figure*}
\center{{
\epsfig{figure=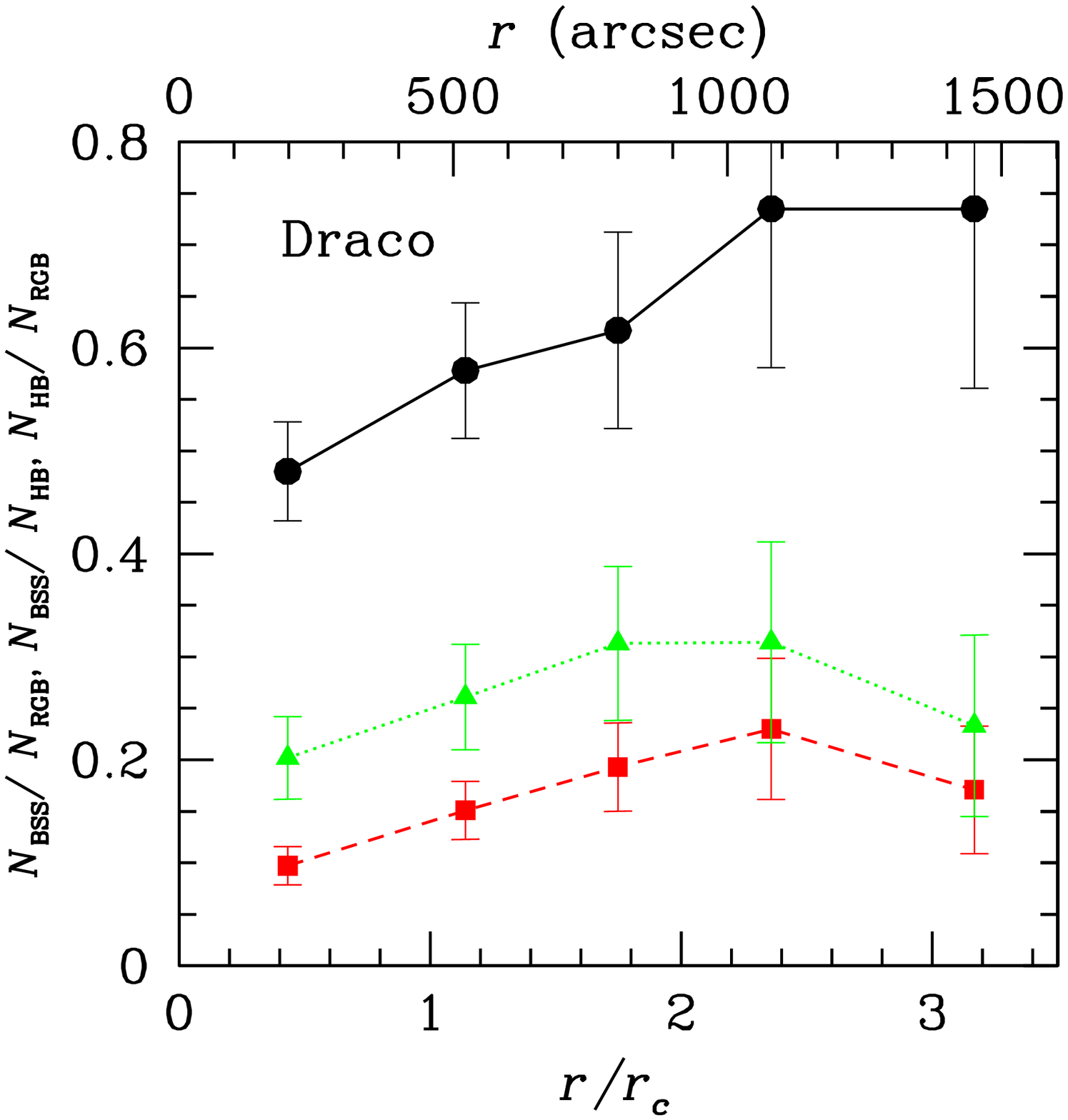,height=8cm}
\epsfig{figure=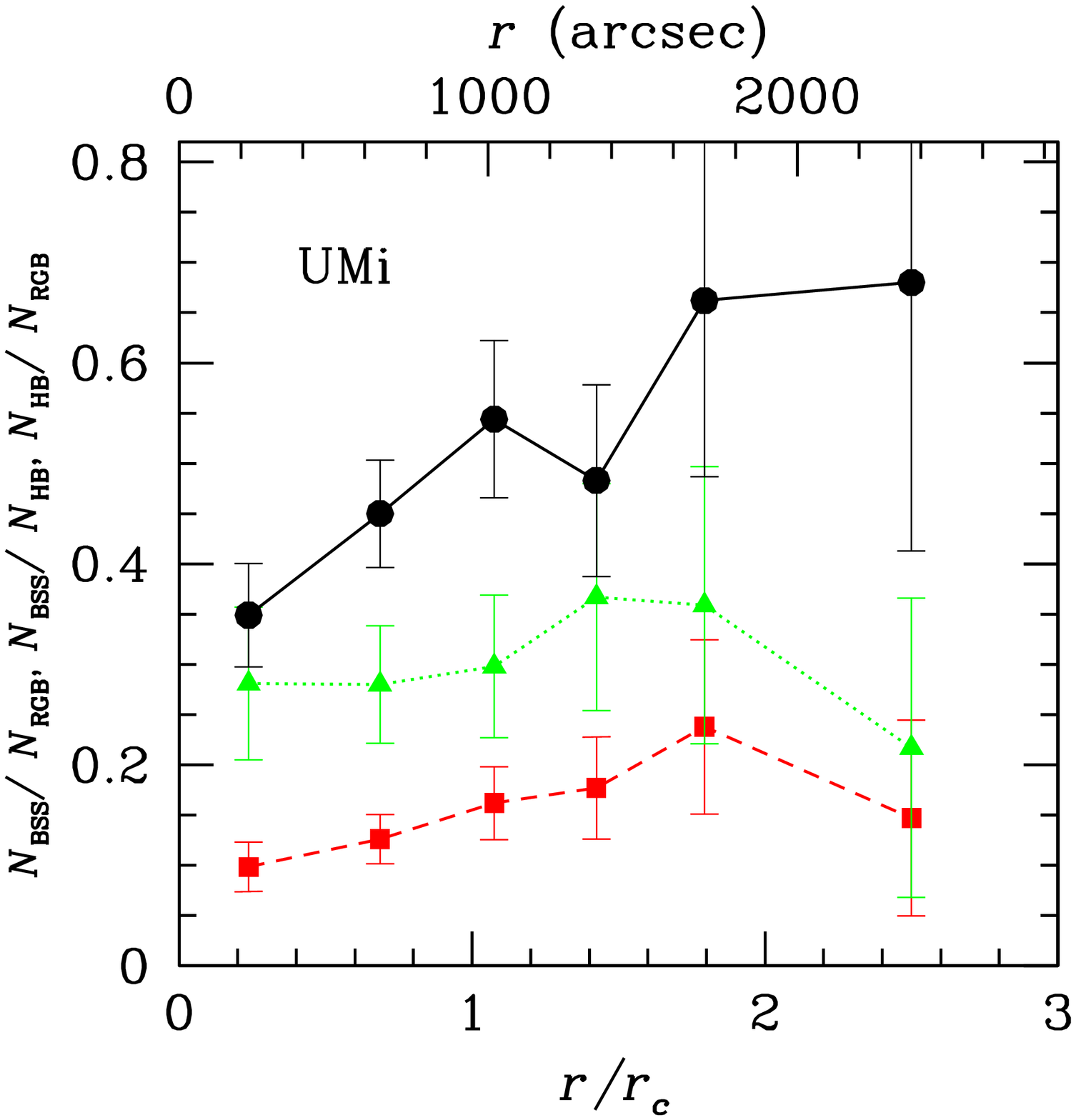,height=8cm}
}}
\caption{\label{fig:fig3}
Observed relative frequency of BSSs normalized to RGB stars (filled squares connected by dashed line) and HB stars (filled triangles connected by dotted line). Observed relative frequency of HB stars normalized to RGB stars (filled circles connected by solid line). Left panel refers to Draco, right panel to Ursa Minor.
}
\end{figure*}
%%%%%%%%%%%%%%%%%%%%%%%%%%%%%%%%%%%%%%%%%%%%%%%%%%%%%%%%%%%%%%%%%%%%%%%%%%%%%%%

\section{Observational properties of BSS candidates}

After accounting for the foreground and background contamination\footnote{Our selection boxes are contaminated both by Milky Way stars in the
foreground, and by extragalactic objects (e.g. quasars and unresolved
galaxies) in the background. Since the foreground component is generally 
dominant,
in the rest of this paper we will refer to any type of contamination as
`foreground', unless the distinction is important.}  in our data,
(see Appendix A for the details), we can extract information about the
radial\footnote{ All the references to `radii' in this paper mean {\it
elliptical radii}. The elliptical radius of a point $(x,y)$ is
$r_{ell}(x,y)^2 = x^2 + [y/(1-e)]^2$, where $e$ is the ellipticity of
the considered galaxy, and the galaxy is assumed to be centred on the
origin, with its major axis aligned with the $x$-axis.}  distribution of
different populations of stars as well as about their luminosity
distribution. Both these quantities are important to understand the
behaviour of BSS candidates [see e.g. M06 for the radial distribution
and Monkman et al. (2006) for the luminosity].

\subsection{Radial distributions}
%%Fig.~\ref{fig:fig2}  shows the surface density profile of BSS,
%%RGB  and HB stars in Draco (left panel) and Ursa Minor (right panel). The
%quantities used in Fig.~2 
%%(in particular the central
%%radius of each annulus and the number of BSS, RGBs and HBs per annulus)
%are listed in Table~2 and Table~3 for Draco and Ursa Minor,
%respectively.  
Fig.~\ref{fig:fig3} shows the radial distribution of the ratio between
the number of BSSs ($N_{{\rm BSS}}$) and that of RGB ($N_{{\rm RGB}}$)
and of HB stars ($N_{{\rm HB}}$). The radial distribution of $N_{{\rm
HB}}$ with respect to $N_{{\rm RGB}}$ is also shown in
Fig.~\ref{fig:fig3}.  The quantities used in Fig.~\ref{fig:fig3} are
listed in Table~2 and Table~3 for Draco and Ursa Minor respectively.

The behaviour of these three relative frequencies
is qualitatively similar in Ursa Minor (right panel) and Draco (left
panel).
%%%%%%%%%%%%%%%%%%%%%%%%%%%%%%% TABLE 2%%%%%%%%%%%%%%%%%%%%%%%%%%%%%%%%%
\begin{table*}
\begin{center}
\caption{Number counts for Draco.}
\leavevmode
\begin{tabular}[!h]{lllllll}
\hline
  $r\,{}$(arcsec)$^{\rm a}$
& $N_{{\rm BSS}}$$^{\rm b}$
& $\epsilon_{{\rm BSS}}$$^{\rm c}$
& $N_{{\rm RGB}}$$^{\rm b}$
& $\epsilon_{{\rm RGB}}$$^{\rm c}$
& $N_{{\rm HB}}$$^{\rm b}$
& $\epsilon_{{\rm HB}}$$^{\rm c}$\\
\hline
 198  & 30.8 (31) & 5.6 & 316 (319)  & 17.9 & 152 (155)  & 12.5\\
 522  & 33.5 (34) & 5.8 & 222 (227)  & 15.1 & 128  (134) & 11.7\\
 800  & 25.1 (26) & 5.1 & 130 (139)  & 11.8 & 80.3 (94)  & 10.1\\
 1080 & 15.1 (16) & 4.0 & 65.3 (75)  &  8.7 & 48.0 (56)  &  7.7\\
 1450 & 11.5 (14) & 3.8 & 67.2 (93) &   9.9 & 49.3 (72)  &  9.2\\
\noalign{\vspace{0.1cm}}
\hline
\end{tabular}
\end{center}
\footnotesize{
$^{\rm {a}}$Centre of the annulus. $^{\rm {b}}$The value out of (in) the parenthesis is after (before) the subtraction of the foreground.  $^{\rm {c}}$ Poissonian error plus a term accounting for foreground subtraction.
}
\end{table*}
%%%%%%%%%%%%%%%%%%%%%%%%%%%%%%%%%%%%%%%%%%%%%%%%%%%%%%%%%%%%%%%%%%%%%%%%%%%%%

%%%%%%%%%%%%%%%%%%%%%%%%%%%%%%% TABLE 3%%%%%%%%%%%%%%%%%%%%%%%%%%%%%%%%%
\begin{table*}
\begin{center}
\caption{Number counts for Ursa Minor.}
 \leavevmode
\begin{tabular}[!h]{lllllll}
\hline
  $r\,{}$(arcsec)$^{\rm a}$
& $N_{{\rm BSS}}$$^{\rm b}$
& $\epsilon_{{\rm BSS}}$$^{\rm c}$
& $N_{{\rm RGB}}$$^{\rm b}$
& $\epsilon_{{\rm RGB}}$$^{\rm c}$
& $N_{{\rm HB}}$$^{\rm b}$
& $\epsilon_{{\rm HB}}$$^{\rm c}$\\
\hline
 225  & 17.9 (18) & 4.2 & 182 (183)  & 13.5 &  63.5 (64)  &   8.0\\
 650  & 29.8 (30) & 5.5 & 237 (239)  & 15.5 & 106.0 (110)  & 10.6\\
 1020 & 23.7 (24) & 4.9 & 146 (149)  & 12.2 &  79.5 (84)  &   9.3\\
 1350 & 14.7 (15) & 3.9 & 82.8 (86)  &  9.3 &  40.0 (41)  &   6.4\\
 1700 &  9.7 (10) & 3.2 & 40.8 (44) &   6.6 &  27.0 (29)  &   5.5\\
 2370 &  2.8 (3)  & 1.7 & 19.1 (21) &   4.6 &  13.0 (15)  &   4.0\\

\noalign{\vspace{0.1cm}}
\hline
\end{tabular}
\end{center}
\footnotesize{
$^{\rm {a}}$Centre of the annulus. $^{\rm {b}}$The value out of (in) the parenthesis is after (before) the subtraction of the foreground.  $^{\rm {c}}$ Poissonian error plus a term accounting for foreground subtraction.
}
\end{table*}
%%%%%%%%%%%%%%%%%%%%%%%%%%%%%%%%%%%%%%%%%%%%%%%%%%%%%%%%%%%%%%%%%%%%%%%%%%%%%

From the shape of the distributions in Fig.~\ref{fig:fig3} we can see that the 
BSS candidates
appear to be slightly less concentrated than both HB and RGB stars. The
relative frequency of BSSs is low especially within 1 $r_c$, 
and there are hints of a maximum at a distance 1.5
$r_c\lesssim{}r\lesssim{}\,{} 2.5\,{}r_c$.  The most remarkable
feature of this distribution is the absence of a central peak in the
relative BSS frequency, which is present in nearly every
globular cluster. 
%The absence of the central peak is crucial to
%understand the nature of these stars.

The distributions of $N_{\rm BSS}/N_{\rm HB}$ and $N_{\rm BSS}/N_{\rm RGB}$ are marginally consistent with flat distributions. In fact, if we try to fit $N_{\rm BSS}/N_{\rm RGB}$ with a flat distribution, the minimum non-reduced $\chi{}^2$ is 8.4 (corresponding to a level of the flat distribution equal to 0.129) and 4.9 (for a level of the  flat distribution equal to 0.130), for Draco and Ursa Minor\footnote{The data points used in the  $\chi{}^2$ analysis are 5 for Draco and 6 for Ursa Minor (see Fig.~\ref{fig:fig3}). There is 1 parameter, i.e. the level of the flat distribution.}, respectively. 
%According to the Kolmogorov-Smirnov (KS) test, the probability that $N_{\rm BSS}/N_{\rm RGB}$ is drawn from a flat distribution is $\sim{}0.7$  for Draco and only $\sim{}0.3$ for Ursa Minor.
The resultant null hypothesis probability for a flat distribution is only $\sim{}0.08$ for Draco and $\sim{}0.43$ for Ursa Minor.

% I would have said a better way to say this is to argue that these results
% imply you cannot reject the flat distribution hypothesis.  I also would
% include the number of degrees of freedom in the text not a footnote and
% quote the reduced chisqu since its easier to interpret.  Why not also
% quote the KS results in the text because they support even more the fact
% that a flat distribution fits the data ?

 Similarly, to fit $N_{\rm BSS}/N_{\rm HB}$ with a flat distribution, the minimum non-reduced $\chi{}^2$ is 2.6 (corresponding to a level of the flat distribution equal to 0.243) and 1.0 (for a level of the  flat distribution equal to 0.288), for Draco and Ursa Minor, respectively. The resultant null hypothesis probability for a flat distribution is $\sim{}0.63$ for Draco and $\sim{}0.96$ for Ursa Minor. 
%With the KS test, the probability that $N_{\rm BSS}/N_{\rm HB}$ is drawn from a flat distribution is only $\sim{}0.04$  for Draco and only $\sim{}0.3$ for Ursa Minor.

%We also checked what is the maximum value of the central bin of $N_{\rm BSS}/N_{\rm RGB}$ ($N_{\rm BSS}/N_{\rm HB}$)

We also checked the probability that $N_{\rm BSS}/N_{\rm RGB}$ and $N_{\rm BSS}/N_{\rm HB}$ are consistent with a distribution rising in the central bin and flat elsewhere\footnote{For this analysis we have 2 parameters: the level of the central bin and the level of the flat distribution for the other bins.}. For Draco, we found that this probability drops below 0.05 if the central bin is a factor of 1.00 (1.49) higher than the outer ones in the case of $N_{\rm BSS}/N_{\rm RGB}$ ($N_{\rm BSS}/N_{\rm HB}$). For Ursa Minor, the probability drops below 0.05 if the central bin is a factor of 1.45 (2.42) higher than the outer ones in the case of $N_{\rm BSS}/N_{\rm RGB}$ ($N_{\rm BSS}/N_{\rm HB}$). Then, we can conclude that the observed distribution of $N_{\rm BSS}/N_{\rm HB}$ and especially $N_{\rm BSS}/N_{\rm RGB}$ are hardly consistent with a central rise like the one we observe in most of globular clusters (M06).

If BSS candidates were young MS stars, we would expect them to be more
concentrated than older stars,
%because the gas is likely to be exhausted later in the centre than
%in the outskirts, and also 
consistent with observations where 
metal-rich (younger) stars 
are typically more centrally concentrated than metal-poor (older) stars
(e.g. Tolstoy et al. 2004, Battaglia et al. 2006).

Furthermore, the spatial distribution of
BSS does not show the clumping which could be expected in the case of a
young population (e.g., Fornax dSph; Battaglia et al. 2006). Thus, the
observed radial distribution of BSS candidates suggests (even if does not prove) that these stars are
BSSs and not a young population.
%%%%%%%%%%%%%%%%%%%%%%%%%%%%%%%%%%% FIGURE 4 %%%%%%%%%%%%%%%%%%%%%%%%%%%%%%%%%%
\begin{figure*}
\center{{ \epsfig{figure=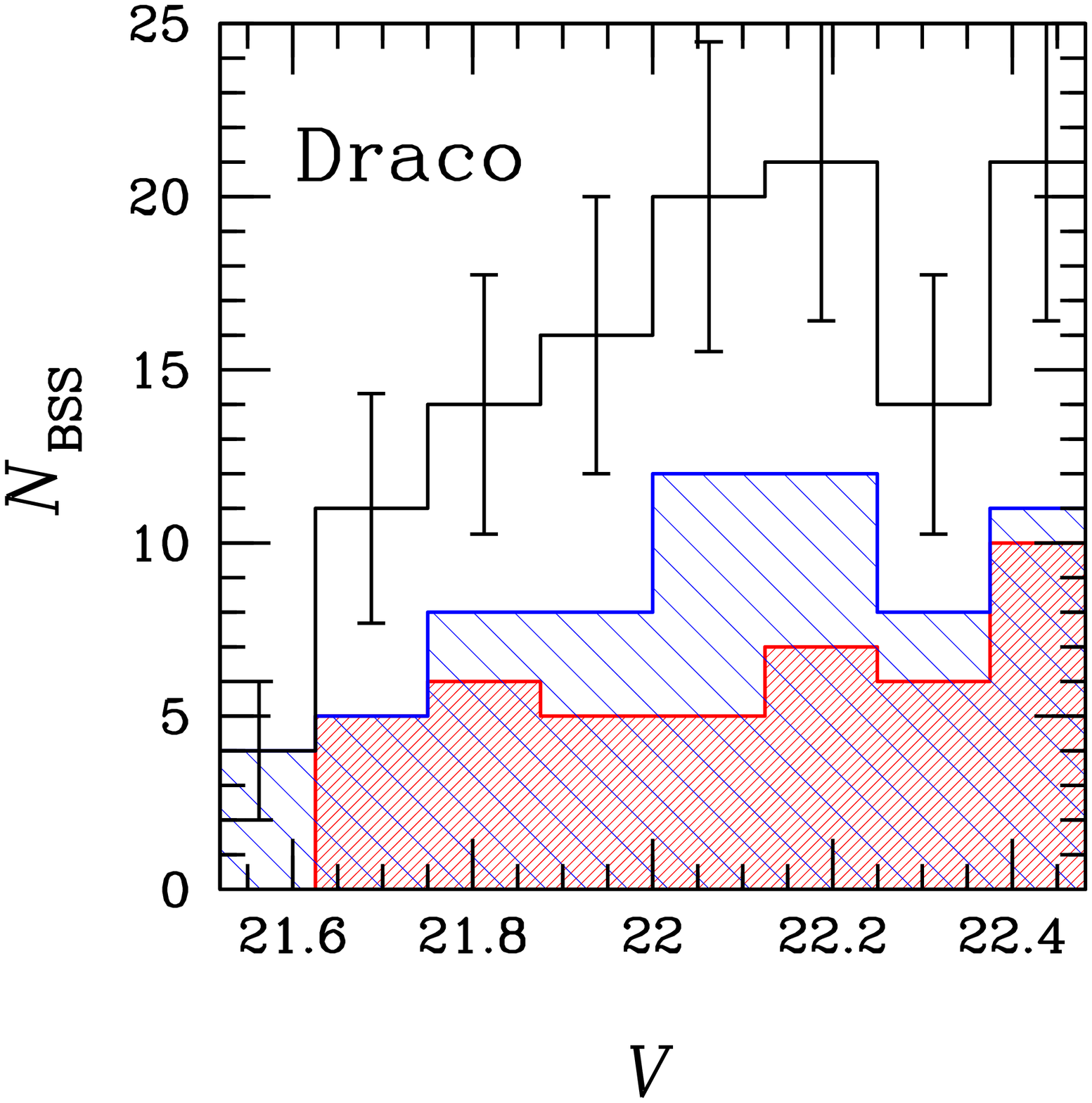,height=8cm}
\epsfig{figure=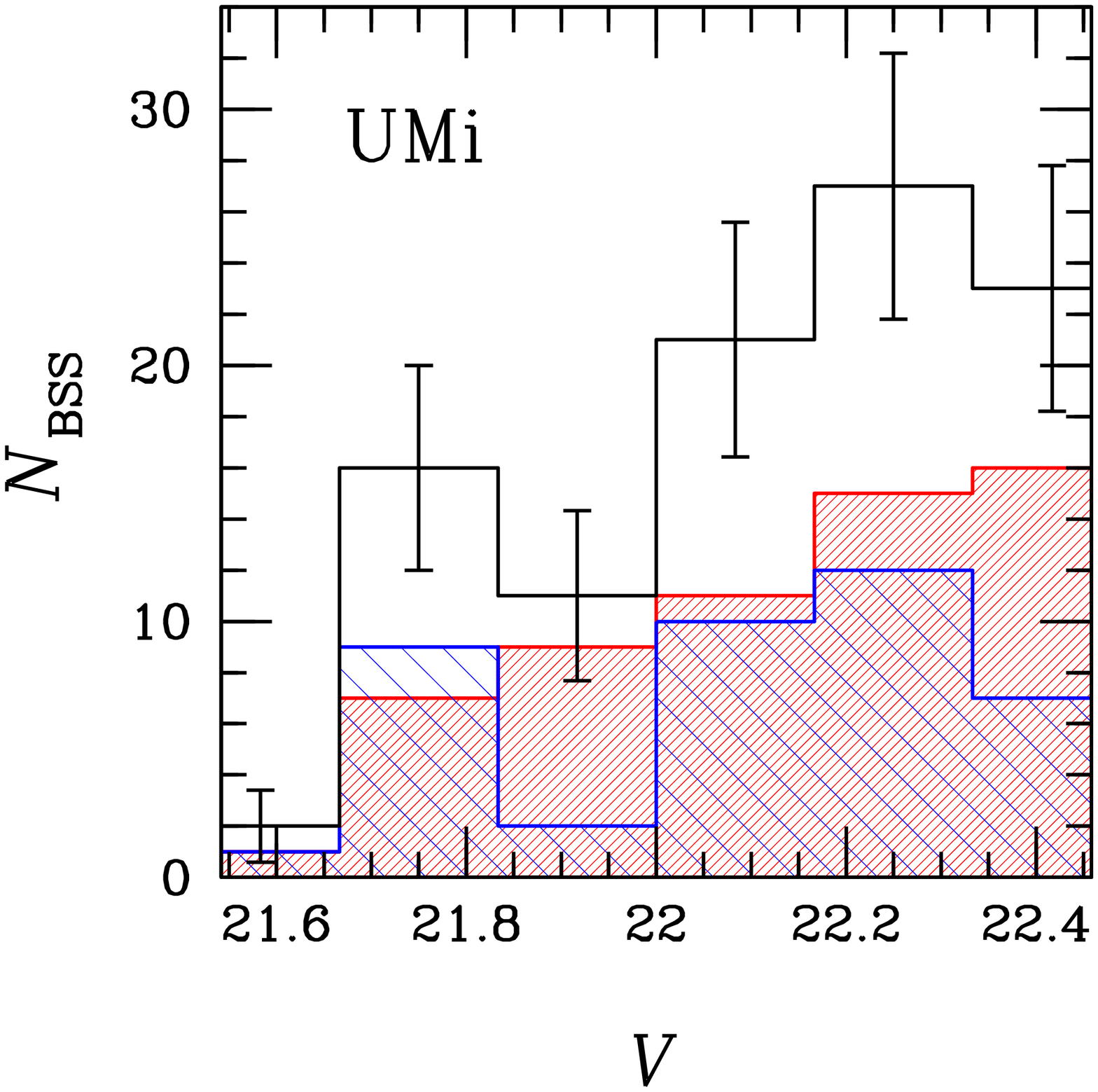,height=8cm} }}
\caption{\label{fig:fig5} Luminosity distribution of BSSs in Draco (left
panel) and Ursa Minor (right panel). The empty histogram represents the
entire sample of BSSs and the error bars  show the Poissonian
error. The lightly hatched (heavily hatched) histogram represents BSSs
with radial position $r>r_c$ ($r<r_c$).  }
\end{figure*}
%%%%%%%%%%%%%%%%%%%%%%%%%%%%%%%%%%%%%%%%%%%%%%%%%%%%%%%%%%%%%%%%%%%%%%%%%%%%%%%

%%%%%%%%%%%%%%%%%%%%%%%%%%%%%%%%%%% FIGURE 5 %%%%%%%%%%%%%%%%%%%%%%%%%%%%%%%%%%
\begin{figure*}
\center{{
\epsfig{figure=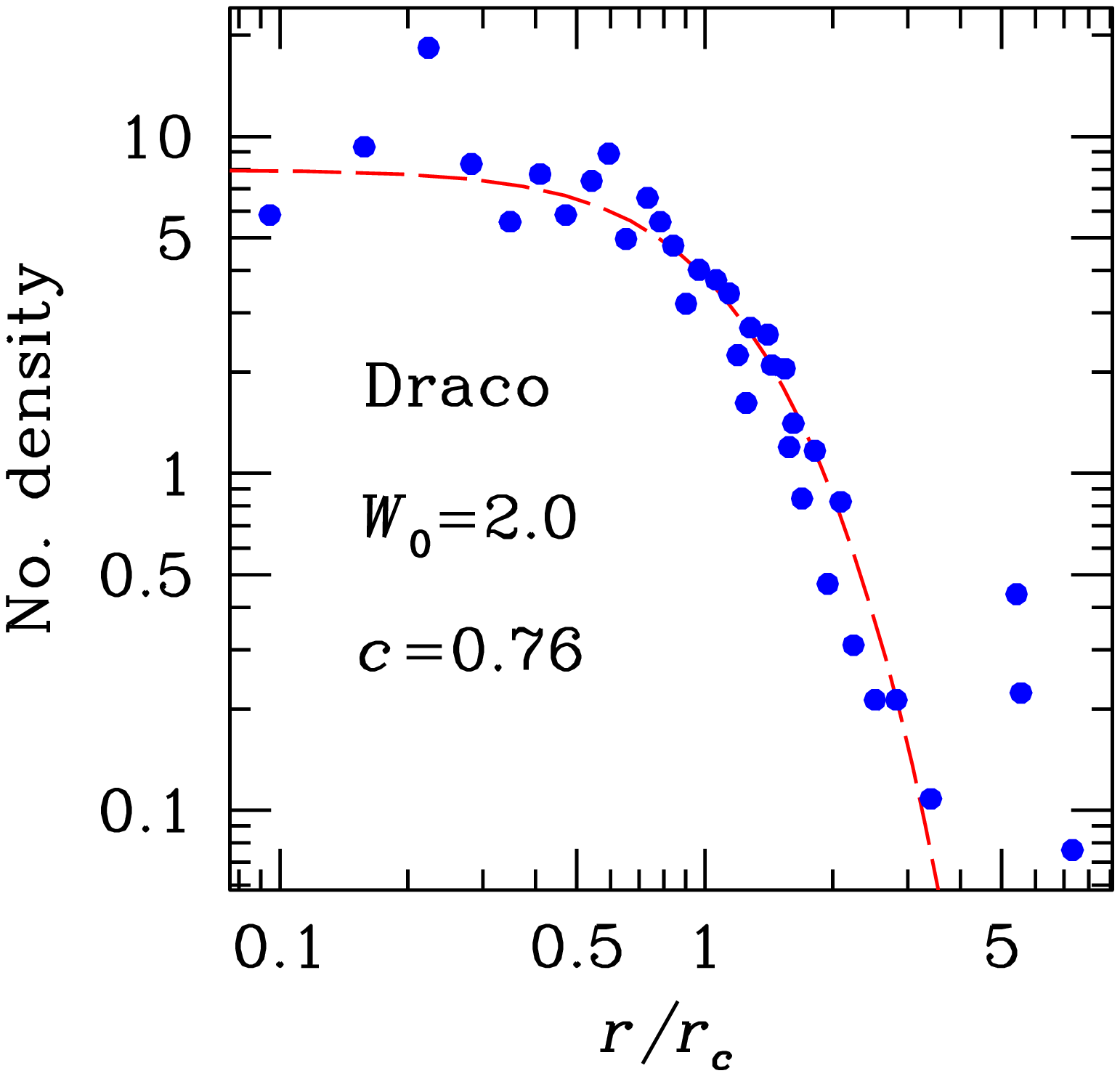,height=8cm}
\epsfig{figure=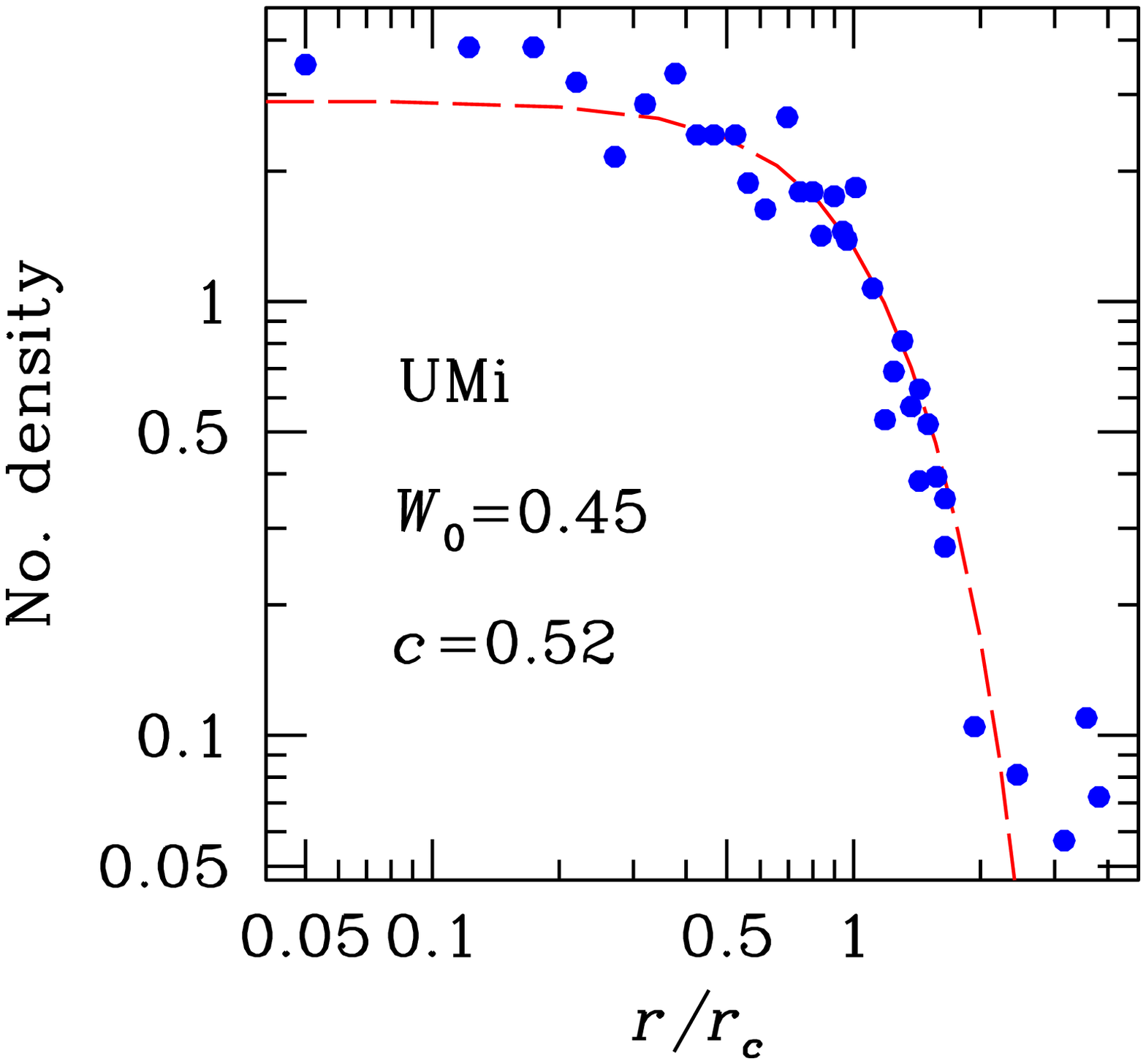,height=8cm}
}}
\caption{\label{fig:fig6} Surface density profile of Draco (left panel)
and Ursa Minor (right panel).  The number density is given in stars per
arcmin$^2$. The filled circles are data points from IH95. The dashed
line is the best-fitting simulation.  }
\end{figure*}
%%%%%%%%%%%%%%%%%%%%%%%%%%%%%%%%%%%%%%%%%%%%%%%%%%%%%%%%%%%%%%%%%%%%%%%%%%%%%%%

%Assuming that these stars are BSSs, 
Are these observed radial distributions
consistent with models of BSS formation and evolution?  According to the
model by M04 and M06, most BSSs in the core of a dense 
host system form from collisions. A central peak in the relative BSS
frequency is expected only if a sufficiently large number of COL-BSSs can
form.

The absence of any central peak in Fig.~\ref{fig:fig3} is consistent with 
this model. In fact, in dSphs the stellar density, even in the
core, is so low that stellar collisions are unlikely to occur\footnote{The collision rate, defined as (Davies, Piotto \& De Angeli 2004) $\Gamma{}\sim{}N_c\,{}n_c\,{}\Sigma{}_{coll}\,{}\sigma{}_c$ (where $\Sigma{}_{coll}$ is the collision cross-section, $N_c$ the number of stars in the core, $n_c$ and $\sigma{}_c$ the stellar number density and velocity dispersion in the core), is more than a factor of $10^5$ smaller in dSphs than in globular clusters.}, and
COL-BSSs cannot form. The
 dearth of BSSs in the centre is also  favoured by
the inefficiency of dynamical friction in dSphs: even if BSSs have higher
mass than both RGB and HB stars, it takes too  long for them to
sink to the centre.

Furthermore, M04 and M06 also predict that MT-BSSs (see Section 4) have a
relative frequency that is almost flat in the centre and slightly rising
in the periphery. This implies that BSS candidates in Draco and Ursa
Minor do behave like MT-BSSs, whereas they are unlikely to be COL-BSSs.

%The BSS candidates in Draco and Ursa Minor are observed to have a relative
%frequency that is almost flat in the centre and slightly rising in the
%periphery, which is consistent with models for MT-BSS (see Section
%4, and M06).

%The presence of such peak  within the core radius is considered
%associated with COL-BSS, which form only in the densest region of the
%host system (M04, M06). Thus, the absence of the central peak seems to exclude the existence of COL-BSS in these dwarf galaxies, as we would have expected, given the low density of these system. 
%{\it Vice versa} the absence of any central peak  in such a low density environment supports the theory that most of BSS in the core of the host system form from collisions. The lack of BSS in the centre is also encouraged by the inefficiency of dynamical friction in dSphs: even if BSS have higher mass than both RGBs and HBs, it takes too much time for them to sink to the centre.

%To check if this consideration holds also for Draco and Ursa Minor, we divided our sample of RGBs (HBs) in blue RGBs (BHBs) and in (RHBs), and we calculated the radial distribution..E' IL CASO DI FARLO?

%L03 also claims that the radial distribution of BSS  inside the
%Sextans dSph depends on their luminosity, i.e. that faint BSS are less
%concentrated than the bright ones. We have checked whether this is true
%also in the case of Draco and Ursa Minor.  

We compared our definitions of BSSs and other populations with those
used by L03 and tested the robustness of different choices.  L03 adopt
a wider definition of BSS, and normalize the frequency of BSSs to the
number of sub-giant branch (SGB) stars. We used both our definition of
BSSs and normalization to the RGB stars and the L03 definition
adapted for the distances and reddenings of Draco/Ursa Minor, combined
with a normalization to SGB stars.
%In Fig.~\ref{fig:fig4} we report the results
%for Draco (left) and Ursa Minor (right panel), adopting the L03
%definition and normalization (but the results do not change if we adopt
%our definitions). 
We do not observe any significant difference.
% between faint ($22.1<V<22.8$) and bright ($20.9<V<22.1$)
%BSS. Indeed, bright BSS in Draco seem slightly less concentrated than
%the faint ones.  The discrepancy between our results and that of L03
%could be explained  with an intrinsic difference of BSS in
%Sextans with respect to Draco and Ursa Minor (see Section 5).

%The distribution of $N_{{\rm BSS}}/N_{{\rm SGB}}$
%%($N_{{\rm SGB}}$ is the number of SGB stars per each annulus) 
%is very
%similar to the distribution of both $N_{{\rm BSS}}/N_{{\rm RGB}}$ and
%$N_{{\rm BSS}}/N_{{\rm HB}}$. These properties suggest 

This suggests that our results
are reasonably independent of the BSS selection
criteria, and also of the stellar population we choose as a normalization
control sample (HB, RGB or SGB).

In Fig.~\ref{fig:fig3} it can also be seen that HB stars also seem to be 
slightly less concentrated than RGB stars.
%, especially in Ursa Minor. 
%The mass
%and the age of these two groups, in fact,  should be similar,
%and the difference in the two spatial distributions is
%puzzling. However, we note that the error bars of $N_{{\rm HB}}/N_{{\rm
%RGB}}$ are quite  large (especially in the case of Ursa
%Minor), so that the distribution is still consistent with a flat one.
This again is consistent with trends seen in other dSph
(Harbeck et al. 2002; 
L03; Tolstoy et al. 2004; Battaglia et al. 2006), suggesting that older 
populations tend to be less centrally concentrated. 
In particular the blue old stellar populations (BHB, blue RGB, etc)
tend to be less concentrated than their red counterparts (RHB, red RGB, etc). 
However, as in the case of $N_{\rm BSS}/N_{\rm HB}$ and $N_{\rm BSS}/N_{\rm RGB}$, $N_{\rm HB}/N_{\rm RGB}$ is also statistically consistent with a flat distribution.
%{\bf [ER: red/blue RGBs???  btw, il Tolstoy et al. 2004 ed il Battaglia et al. 2006 confermano la cosa non solo attraverso la fotometria ma anche dagli spettri]}

%%%%%%%%%%%%%%%%%%%%%%%%%%%%%%%%%%% FIGURE 4 %%%%%%%%%%%%%%%%%%%%%%%%%%%%%%%%%%
%\begin{figure*}
%\center{{
%\epsfig{figure=lee.eps,height=8cm}
%\epsfig{figure=lee_ursa_nofg.eps,height=8cm}
%}}
%\caption{\label{fig:fig4}
%Observed relative frequency of BSS normalized to SGBs. Left panel refers to Draco, right panel to Ursa Minor. The solid line refers to all BSS, the dotted line to faint BSS and the dashed line to bright BSS. BSS are defined [accordingly to L03] as stars with $20.9<$V$<22.8$ and with $-0.1<$(V-I)$<0.2$ for Draco and $0.1<$(V-I)$<0.4$ for Ursa Minor. SGBs are RGBs with $20.9<$V$<22.8$. Faint (bright) BSS are BSS with $22.1<$V$<22.8$ ($20.9<$V$<22.1$).
%}
%\end{figure*}
%%%%%%%%%%%%%%%%%%%%%%%%%%%%%%%%%%%%%%%%%%%%%%%%%%%%%%%%%%%%%%%%%%%%%%%%%%%%%%%

\subsection{Luminosity distribution}
The luminosity distribution is another important indicator of BSS
properties. Recent papers (Ferraro et al. 2003; Monkman et al. 2006)
 suggest a correlation between the brightness of BSSs and their
radial distance from the the centre of a
globular cluster. Bright BSSs tend to be
more concentrated than the faint ones. In
turn, if the model of M04 and M06 is correct, such a correlation suggests
that the centrally concentrated COL-BSSs tend to be brighter than
MT-BSSs. This  scenario makes sense, as
COL-BSSs should conserve a large fraction of the mass of the colliding
progenitors (Benz \& Hills 1987, 1992; Sills et al. 2001; Freitag \&
Benz 2005), whereas the mass-transfer process is not as efficient (Pols
\& Marinus 1994; Tian et al. 2006).

In dSphs, where only MT-BSSs are likely to form, we do not expect to see
a correlation between the brightness of BSSs and their radial
position. This prediction is completely supported by the observed BSS
luminosity distribution of Draco (Fig.~\ref{fig:fig5}, left panel) and
Ursa Minor (Fig.~\ref{fig:fig5}, right panel). The open histograms in
Fig.~\ref{fig:fig5} show the total luminosity distribution, while the
light and heavy hatched histograms show the luminosity distribution of
BSSs which are located outside and within $r_c$, respectively.  According 
to the Kolmogorov-Smirnov (KS)  test, the probability that light and heavy hatched histograms are drawn from the same distributions is $>0.999$
%9994 and 0.9999565 
both for Draco and for Ursa Minor. Thus, there is
% couldn't we simply say < 1.0e-6 and < 1.0e-4 respectively
no statistically significant difference between these distributions,
indicating no dependence of the luminosity function on the radial
distance. This fact is at odds with the findings by L03, who observe a
correlation between the brightness of BSSs in Sextans and their radial
distance. Since (as we checked) our luminosity distributions do not
change by adopting the BSS selection criteria by L03, we suggest that
this is due to an intrinsic difference between Sextans and Draco/Ursa
Minor (see Section 5).

%In order to be sure that our selection criteria do not affect significantly our results, we made also some checks by adopting different criteria. For example, for comparison with L03, we considered Draco BSS all those stars with V=[ 20.9, 22.8] and (V-I)=[-0.3, 0.3]

\section{The simulations}

The data presented in the previous sections
show that BSS candidates in Draco and Ursa Minor behave like
MT-BSSs. As a further check, we ran for Draco and
Ursa Minor the same kind of dynamical simulations that were performed by
M04 and M06 for BSSs in globular clusters.

\subsection{Method}
We adopt the upgraded version of the code by Sigurdsson \& Phinney
(1995) already described in M04 and M06. The code integrates the
dynamics of BSSs, under the influence of the galactic potential, of
dynamical friction (using Chandrasekhar formula) and of distant
encounters with other stars. Also three-body encounters are implemented in
the code; but they are unimportant in the runs for dSphs. 
%{\bf We expect that for dSphs also dynamical friction and distant encounters are much less efficient than in globular clusters}.

The potential of the host galaxy is represented by a time independent
multimass King model. The classes of mass are the same as in M04, and
the assumed turn-off mass is 0.8 $M_\odot{}$. To calculate the
potential, we input the observed core density ($n_c$) and velocity
dispersion ($\sigma{}_c$) of Draco and Ursa Minor (the adopted values
are  listed in Table~1), and we modify the value of the
central adimensional potential, $W_0$ (defined in Sigurdsson \& Phinney
1995), until we reproduce the concentration and the
density profile of the galaxy under consideration. 
In Fig.~\ref{fig:fig6}
 the density profiles
of the best-fitting  King models are compared with the data of
IH95. As expected, the best-fitting
value of $W_0$ is a factor 5$-$20 lower than the common values assumed in
globular clusters.

BSSs are generated with a given position, velocity and mass. Initial
positions are randomly chosen according to a probability distribution
homogeneous in the radial distance from the centre. This means that BSSs
are initially distributed according to an isothermal sphere, as we
expect for MT-BSSs (see M04, M06). The minimum and the maximum value of
the distribution of initial radial distances, $r_{min}$ and $r_{max}$,
have been tuned in order to find the best-fitting simulation (Table 4
and 5 report the most significant runs and their parameters for Draco
and Ursa Minor, respectively).

%%%%%%%%%%%%%%%%%%%%%%%%%%%%%%% TABLE 4%%%%%%%%%%%%%%%%%%%%%%%%%%%%%%%%%
\begin{table*}
\begin{center}
\caption{Simulation parameters and $\chi{}^2$ for Draco.}
\leavevmode
\begin{tabular}[!h]{llllllll}
\hline
Run 
&  $r_{min}/r_c$
& $r_{max}/r_c$
& $v_{kick}/\sigma{}_c$ 
& $t_{last}$ (Gyr)
& $m_{\rm BSS}$ ($M_\odot{}$)
& $\chi{}_{\rm RGB}^2$
& $\chi{}_{\rm HB}^2$\\
\hline
 A1       & 0.8 & 3.5 & 0 & 2 & 1.3  &  0.31 & 0.26 \\
 A2       & 0.8 & 3.5 & 0 & 1 & 1.3  &  0.54 & 0.47 \\ 
 A3       & 0.8 & 3.5 & 0 & 4 & 1.3  &  0.58 & 0.49 \\
 A4       & 0.8 & 3.5 & 0 & 10 & 1.3 &  0.64 & 0.54 \\
 B1       & 0.0 & 3.5 & 0 & 2 & 1.3  &  4.51 & 4.08 \\
 B2       & 0.2 & 3.5 & 0 & 2 & 1.3  &  2.46 & 2.23 \\
 B3       & 0.5 & 3.5 & 0 & 2 & 1.3  &  0.62 & 0.60 \\
 B4       & 1.0 & 3.5 & 0 & 2 & 1.3  &  1.09 & 0.94 \\ 
 C1       & 0.8 & 4.5 & 0 & 2 & 1.3  &  4.02 & 3.64 \\  
 C2       & 0.8 & 3.0 & 0 & 2 & 1.3  &  1.89 & 1.72 \\ 
 C3       & 0.8 & 2.5 & 0 & 2 & 1.3  &  7.03 & 6.47 \\
 D1       & 0.0 & 4.5 & 0 & 2 & 1.3  &  2.74 & 2.47 \\
 D2       & 0.0 & 3.0 & 0 & 2 & 1.3  &  8.67 & 7.89 \\
 E1       & 0.8 & 3.5 & 0 & 2 & 1.1  &  0.31 & 0.26 \\ 
 E2       & 0.8 & 3.5 & 0 & 2 & 1.5  &  0.56 & 0.47 \\ 
 F1       & 0.8 & 3.5 & 1.& 2 & 1.3  &  0.85 & 0.73 \\

\noalign{\vspace{0.1cm}}
\hline
\end{tabular}
\end{center}
\footnotesize{
$\chi{}_{\rm RGB}^2$ ($\chi{}_{\rm HB}^2$) indicates the $\chi{}^2$ of the $N_{{\rm BSS}}/N_{{\rm RGB}}$ ($N_{{\rm BSS}}/N_{{\rm HB}}$) distribution. The reported values of $\chi{}_{\rm RGB}^2$ and $\chi{}_{\rm HB}^2$ are not reduced and have been calculated on the basis of 5 data points.
}
\end{table*}
%%%%%%%%%%%%%%%%%%%%%%%%%%%%%%%%%%%%%%%%%%%%%%%%%%%%%%%%%%%%%%%%%%%%%%%%%%%%%

%%%%%%%%%%%%%%%%%%%%%%%%%%%%%%% TABLE 5%%%%%%%%%%%%%%%%%%%%%%%%%%%%%%%%%
\begin{table*}
\begin{center}
\caption{Simulation parameters and $\chi{}^2$ for Ursa Minor.}
\leavevmode
\begin{tabular}[!h]{llllllll}
\hline
Run 
&  $r_{min}/r_c$
& $r_{max}/r_c$
& $v_{kick}/\sigma{}_c$ 
& $t_{last}$ (Gyr)
& $m_{\rm BSS}$  ($M_\odot{}$)
& $\chi{}_{\rm RGB}^2$
& $\chi{}_{\rm HB}^2$\\
\hline
 A1  & 0.5 & 1.9  & 0 & 2  & 1.3  & 0.42  & 0.36  \\
 A2  & 0.5 & 1.9  & 0 & 1  & 1.3  & 0.32  & 0.28  \\
 A3  & 0.5 & 1.9  & 0 & 4  & 1.3  & 0.20  & 0.17  \\
 A4  & 0.5 & 1.9  & 0 & 10 & 1.3  & 0.34  & 0.29  \\
 B1  & 0.0 & 1.9  & 0 & 2  & 1.3  & 1.68  & 1.50  \\
 B2  & 0.2 & 1.9  & 0 & 2  & 1.3  & 0.58  & 0.53  \\ 
 B3  & 0.8 & 1.9  & 0 & 2  & 1.3  & 1.93  & 1.61  \\
 B4  & 1.0 & 1.9  & 0 & 2  & 1.3  & 3.51  & 2.95  \\
 C1  & 0.5 & 3.0  & 0 & 2  & 1.3  & 18.07 & 16.06 \\
 C2  & 0.5 & 2.5  & 0 & 2  & 1.3  &  4.79 &  4.24 \\ 
 C3  & 0.5 & 1.0  & 0 & 2  & 1.3  &  6.99 &  6.21 \\ 
 D1  & 0.0 & 3.0  & 0 & 2  & 1.3  & 12.60 & 11.29 \\
 D2  & 0.0 & 1.5  & 0 & 2  & 1.3  &  5.53 &  4.89 \\
 E1  & 0.5 & 1.9  & 0 & 2  & 1.1  &  0.42 &  0.36 \\
 E2  & 0.5 & 1.9  & 0 & 2  & 1.5  &  0.43 &  0.36 \\
 F1  & 0.5 & 1.9  & 1 & 2  & 1.3  &  2.67  & 2.33 \\
\noalign{\vspace{0.1cm}}
\hline
\end{tabular}
\end{center}
\footnotesize{
$\chi{}_{\rm RGB}^2$ ($\chi{}_{\rm HB}^2$) indicates the $\chi{}^2$ of the $N_{{\rm BSS}}/N_{{\rm RGB}}$ ($N_{{\rm BSS}}/N_{{\rm HB}}$) distribution. The reported values of $\chi{}_{\rm RGB}^2$ and $\chi{}_{\rm HB}^2$ are not reduced and have been calculated on the basis of 6 data points.
}
\end{table*}
%%%%%%%%%%%%%%%%%%%%%%%%%%%%%%%%%%%%%%%%%%%%%%%%%%%%%%%%%%%%%%%%%%%%%%%%%%%%%

Initial velocities are generated from the distributions described in
Sigurdsson \& Phinney (1995). In  most runs, no initial kicks
 are given to BSSs,  because  they are expected to be
MT-BSSs. We also made some (physically unrealistic) check run, were a
kick velocity ($v_{kick}$) is given to BSSs born inside the core.

 In most of the runs the mass of the BSSs is assumed to be $m_{\rm
BSS}=1.3\,{}M_\odot{}$. We made check runs with masses in the range from
1.1 to 1.5 $M_\odot{}$ (higher masses are unlikely, at least for some
globular clusters; see Ferraro et al. 2004, 2006). This range of masses
is also consistent with the isochrones for our data of Draco and Ursa
Minor (see Appendix B).
%, without noting substantial differences. CONTROLLARE DI NUOVO!

Each BSS is evolved for a time $t$, randomly selected  from a
homogeneous distribution between $t=0$ and $t=t_{last}$. The parameter
$t_{last}$ represents the lifetime of BSSs (see M04, M06). We made runs
with $t_{last}$=1, 2, 4, 10 Gyr.
%The results are quite similar, since dynamical friction is quite
%inefficient in dSphs. Thus, we cannot estimate the age of BSS in dSphs
%merely on the basis of dynamics. CONTROLLARE DI NUOVO!!!!

%dN(r)\propto{}r^{-2}

\subsection{Comparison with observations}
%%%%%%%%%%%%%%%%%%%%%%%%%%%%%%%%%%% FIGURE 6 %%%%%%%%%%%%%%%%%%%%%%%%%%%%%%%%%%
\begin{figure*}
\center{{
\epsfig{figure=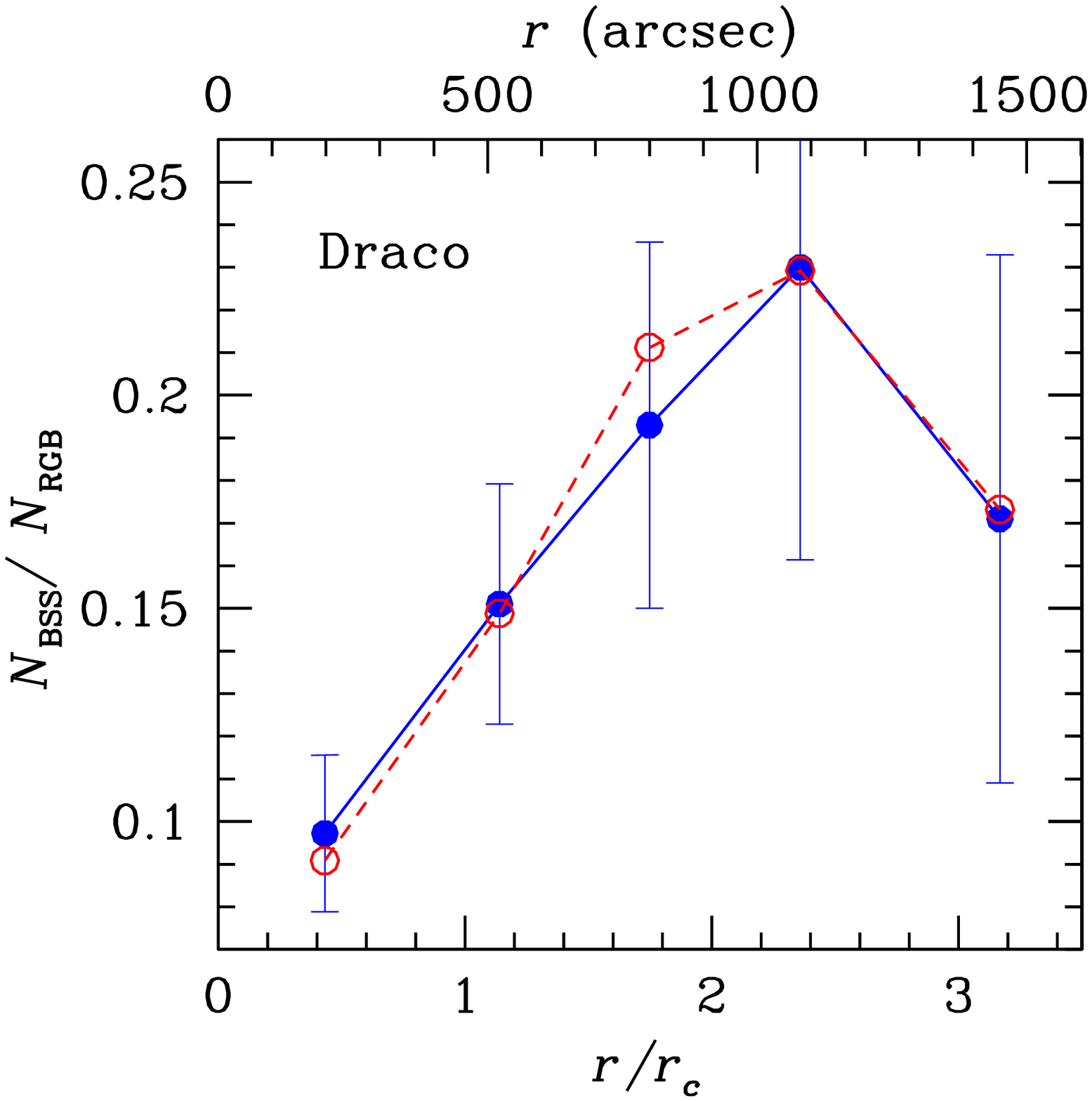,height=8cm}
\epsfig{figure=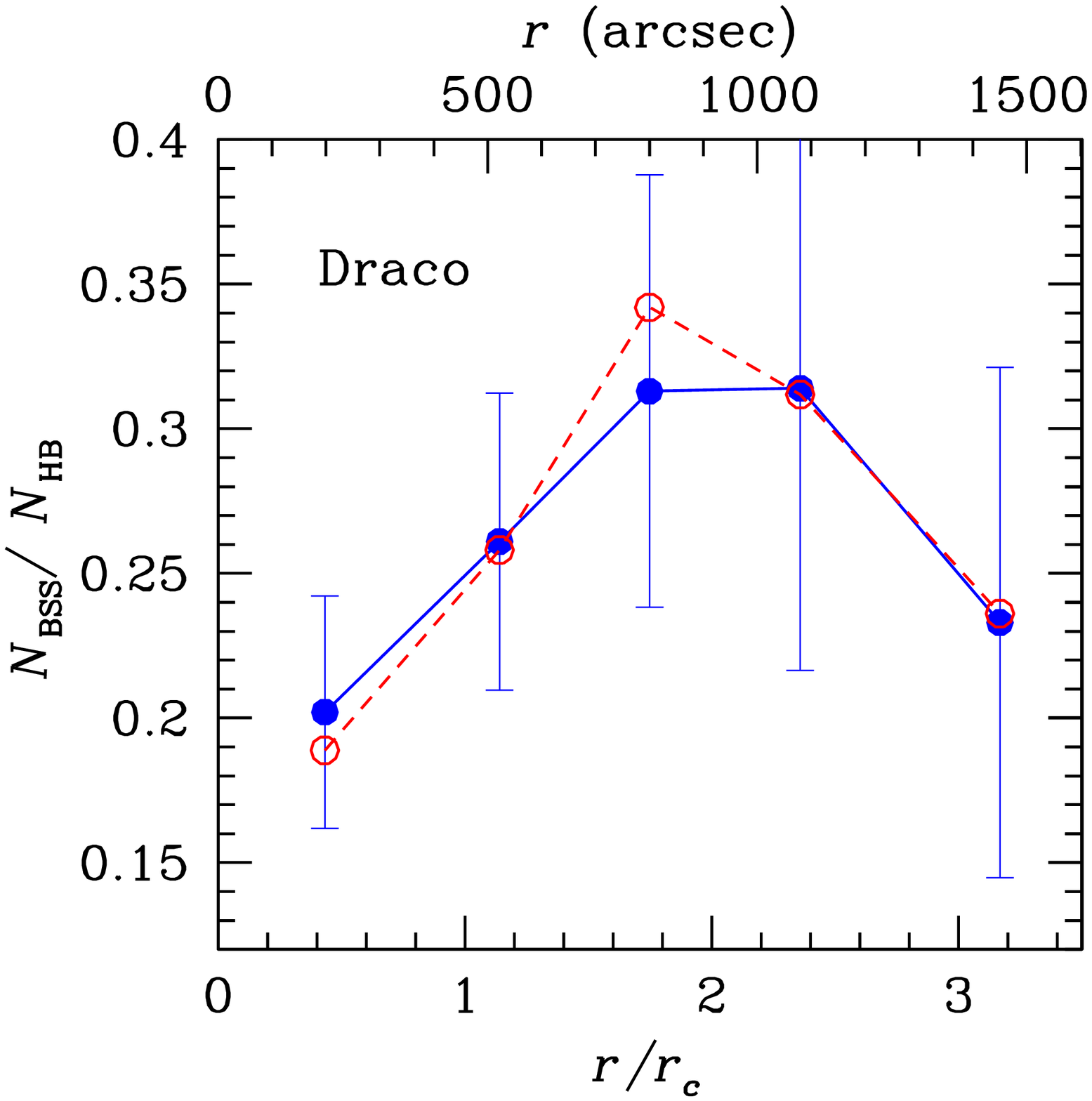,height=8cm}
}}
\caption{\label{fig:fig7}
Relative frequency of BSSs normalized to RGB  (left panel) and HB stars (right panel) in Draco. The filled circles connected by the solid line are the measurements (the same as in Fig.~\ref{fig:fig3}). The open circles connected by the dashed line are the fiducial model (run A1).
}
\end{figure*}
%%%%%%%%%%%%%%%%%%%%%%%%%%%%%%%%%%%%%%%%%%%%%%%%%%%%%%%%%%%%%%%%%%%%%%%%%%%%%%%
%%%%%%%%%%%%%%%%%%%%%%%%%%%%%%%%%%% FIGURE 7 %%%%%%%%%%%%%%%%%%%%%%%%%%%%%%%%%%
\begin{figure*}
\center{{
\epsfig{figure=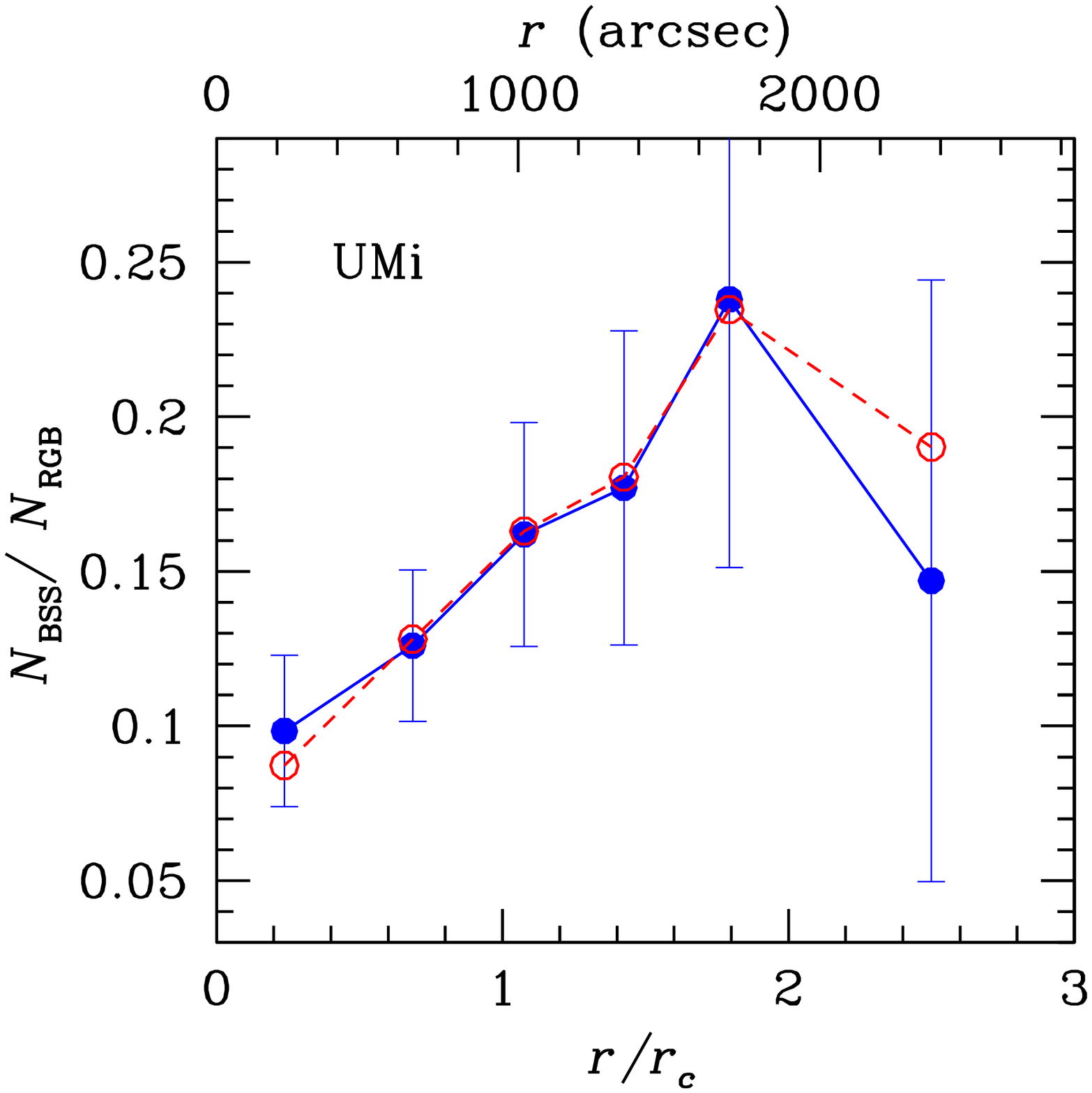,height=8cm}
\epsfig{figure=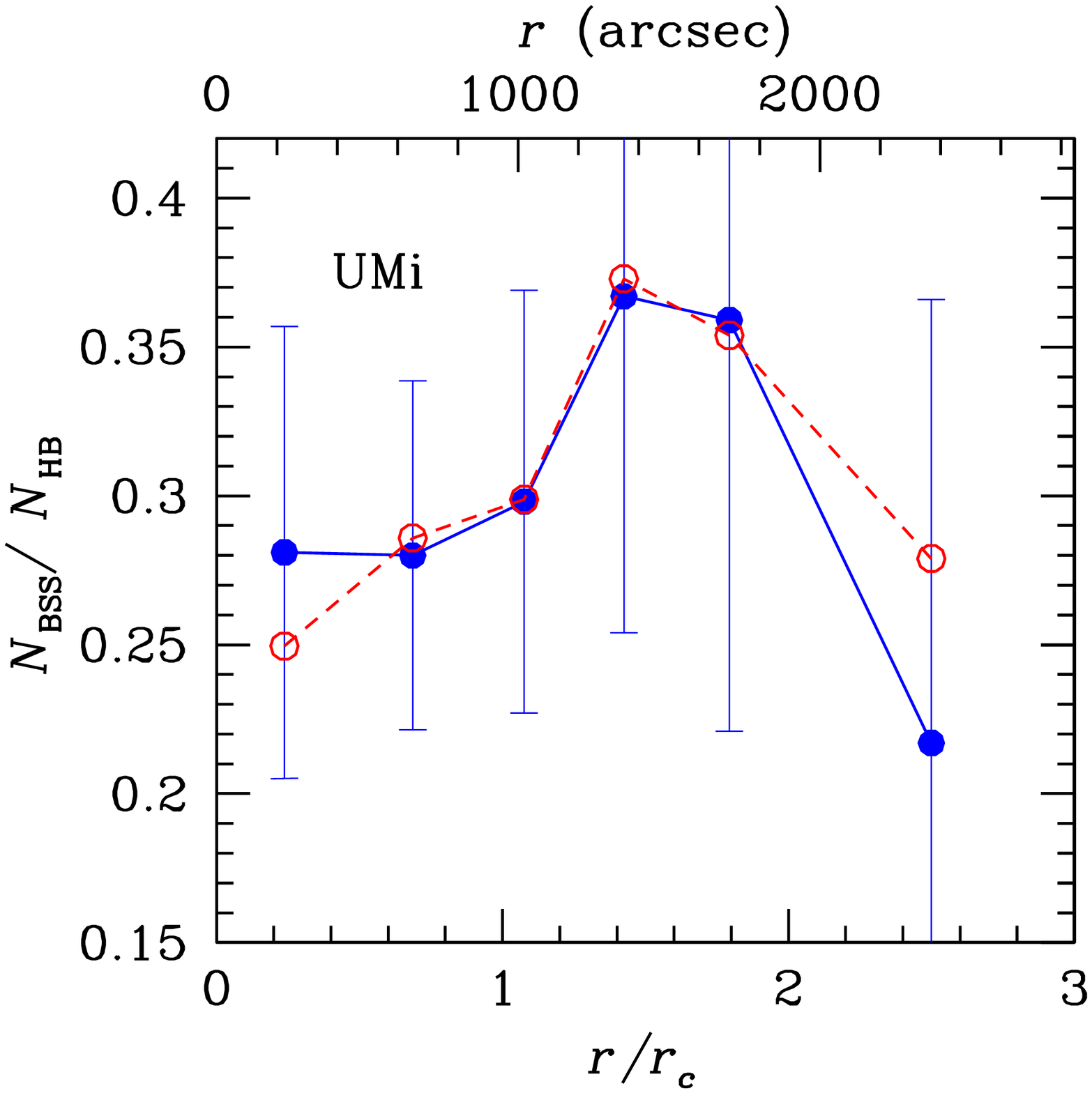,height=8cm}
}}
\caption{\label{fig:fig8}
Relative frequency of BSSs normalized to RGB  (left panel) and HB stars (right panel) in Ursa Minor. The filled circles connected by the solid line are the measurements (the same as in Fig.~\ref{fig:fig3}). The open circles connected by the dashed line are the fiducial model (run A1).
}
\end{figure*}
%%%%%%%%%%%%%%%%%%%%%%%%%%%%%%%%%%%%%%%%%%%%%%%%%%%%%%%%%%%%%%%%%%%%%%%%%%%%%%%

We ran different simulations, adopting different masses and lifetimes
for BSSs, and varying the interval [$r_{min}$, $r_{max}$] where MT-BSSs
are allowed to form.  For each of them, we obtain the
distributions $N_{\rm BSS}/N_{\rm RGB}$ and $N_{\rm BSS}/N_{\rm HB}$,
and calculate their $\chi{}^2$ ($\chi{}_{\rm RGB}^2$ and $\chi{}_{\rm
HB}^2$, respectively) with respect to observations.

From the $\chi{}^2$ analysis (Tables 4 and 5) it appears that the
lifetime of BSSs, $t_{last}$, does not affect the results: runs A1, A2,
A3 and A4, which differ only for $t_{last}$,
%($t_{last}$=1, 2, 4 and 10 Gyr, respectively),
have $\chi{}_{\rm RGB}^2\sim{}\chi{}_{\rm
HB}^2\sim{}1$, both in Draco and in Ursa Minor. Then, in the case of
dSphs, our simulations cannot constrain the age of BSSs. The reason
is that dSphs are dynamically 'quiet' environments, where, due to the low
density, both dynamical friction and close interactions are inefficient.

On the other hand, if BSSs burn a tiny amount of hydrogen, acquired from
the companion stars, they are expected to be relatively short lived.
Thus, the run with $t_{last}=10$ Gyr (A4 for both Draco and Ursa Minor)
is likely unrealistic. In the following, we will consider $t_{last}=2$
Gyr as the fiducial value, for analogy with the findings of M04 and M06
for globular clusters, and because an age of about $2$ Gyr is suggested
also by isochrones (see Appendix~B)
%Ferraro et
%al. 1993, 1997 {\bf [MM: MA QUI SAREBBE MEGLIO SE CI FACESSIMO DELLE
%ISOCHRONE DI NOSTRO PUGNO] [ER: ??? ne parliamo a voce] [MM: devi guardare dove va a finire la massa di turn off dell'isocrona che fitta meglio le BSS assumendo che siano MS: questa ti da' all'incirca la massa]}).

Also the mass of the BSS is not a crucial parameter: runs A1, E1 and E2,
which differ only in the BSS mass,
%($m_{\rm BSS}=$ 1.3, 1.1 and 1.5 $M_\odot{}$, respectively),
have $\chi{}_{\rm RGB}^2\sim{}\chi{}_{\rm
HB}^2\lesssim{}1$. In most of the runs, we assume as fiducial value $m_{\rm
BSS}=1.3\,{}M_\odot{}$.  Masses larger than $m_{\rm
BSS}\sim{}1.4-1.5\,{}M_\odot{}$ tend to be discarded by observations,
both in our data (see Appendix~B) and in globular clusters (Ferraro et
al. 2006).

The parameters which mainly affect our results are the lower and upper
limit of the initial radial position distribution ($r_{min}$ and
$r_{max}$). We remind that initial positions in such simulations
represent the point where a binary which is undergoing mass transfer
turns into a BSS.

The best-fitting value for $r_{min}$ is similar for Draco and Ursa
Minor, and is equal to 0.8 $r_c$ and 0.5 $r_c$, respectively. All the
values of $r_{min}$ from 0 to $\sim{}1\,{}r_c$ give acceptable
$\chi{}^2$ (see e.g. runs B1$-$B4 both for Draco and for Ursa Minor).

The best-fitting $r_{max}$ (expressed in terms of $r_c$) is a factor of
$\sim{}2$ larger for Draco (3.5 $r_c$) than for Ursa Minor (1.9
$r_c$). Indeed, it is possible to reproduce Draco BSSs, recovering an
acceptable $\chi{}^2$, also with $2.5\leq{}r_{max}/r_c\leq{}4.5$,
whereas in the case of Ursa Minor $r_{max}>2.5\,{}r_c$ and
$r_{max}<\,{}r_c$ are inconsistent with observations (see e.g. runs
C1$-$C3 both for Draco and for Ursa Minor).

This discrepancy might be due only to the different normalization. In
fact, the core radius of Ursa Minor is approximately twice as large as
that of Draco. In physical units, the best fits are
$r_{min}\sim{}370$ and $r_{max}\sim{}1600$ arcsec (150 and 640 pc) for
Draco, and $r_{min}\sim{}470$ and $r_{max}\sim{}1800$ arcsec (160 and
620 pc) for Ursa Minor.

%Furthermore, the real
%extension of the tidal radius of Ursa Minor and especially of Draco is
%still under debate. For $r_t$ we adopt the conservative values by IH95;
% but there are claims that the tidal radius of
%Draco is a factor $\sim{}1.5$  larger (A01;
%Piatek et al. 2001).

%%%%%%%%%%%%%%%%%%%%%%%%%%%%%%%%%%% FIGURE 4 %%%%%%%%%%%%%%%%%%%%%%%%%%%%%%%%%%
%\begin{figure*}
%\center{{
%\epsfig{figure=lee.eps,height=8cm}
%\epsfig{figure=lee_ursa_nofg.eps,height=8cm}
%}}
%\caption{\label{fig:figlee}
%Observed relative frequency of BSS normalized to SGB stars. Left panel refers to Draco, right panel to Ursa Minor. The solid line refers to all BSS, the dotted line to faint BSS and the dashed line to bright BSS. BSS are defined [accordingly to L03] as stars with $20.9<$V$<22.8$ and with $-0.1<$(V-I)$<0.2$ for Draco and $0.1<$(V-I)$<0.4$ for Ursa Minor. SGB are RGB stars with $20.9<$V$<22.8$. Faint (bright) BSS are BSS with $22.1<$V$<22.8$ ($20.9<$V$<22.1$).
%}
%\end{figure*}
%%%%%%%%%%%%%%%%%%%%%%%%%%%%%%%%%%%%%%%%%%%%%%%%%%%%%%%%%%%%%%%%%%%%%%%%%%%%%%%

The runs labelled as F1 in the case of both Draco and Ursa Minor have
been set up by taking the best-fitting parameters (runs labelled as A1)
and adding a small kick velocity $v_{kick}=\sigma{}_c$ to BSSs born inside $r_c$. This check is
physically unrealistic, as the natal kick is associated with COL-BSSs,
which cannot form in dSphs. Interestingly, the $\chi{}^2$ is quite good
in both cases. However, we note that more than 10 per cent of BSSs are
'spuriously' ejected in these runs.

Figs.~\ref{fig:fig7} and \ref{fig:fig8}  compare our fiducial
model (run A1) with observations, for Draco and Ursa Minor,
respectively. The good agreement with data is evident: the model has
$\chi{}^2_{RGB}=\chi{}^2_{HB}\sim{}0.3$ for Draco (5 data points) and
$\chi{}^2_{RGB}=\chi{}^2_{HB}\sim{}0.4$ for Ursa Minor (6 data points).
%In Figs.~7 and 8 we report also the comparison with a flat distribution of $N_{\rm BSS}/N_{\rm RGB}$ and $N_{\rm BSS}/N_{\rm HB}$.

 In summary, the dynamical simulations reproduce the
observations very well
for all the possible MT-BSS masses and lifetimes in the
range allowed by the models. The best fit is achieved for the model with
$r_{min}=0.8\,{}r_c$ and $r_{max}=3.5\,{}r_c$ for Draco, and with
$r_{max}=0.5\,{}r_c$ and $r_{max}=1.9\,{}r_c$ for  Ursa Minor. 
However, all $r_{min}$ from 0 to $r_c$ are acceptable, as
well as all the $r_{max}$ within $\approx{}0.5\,{}r_c$ from the
best-fitting value. Thus, BSS candidates are consistent with a
population initially distributed in an
isothermal sphere between the centre of the galaxy and the tidal
radius. This result agrees with the model of BSS formation from
mass-transferring binaries, and hints that BSS candidates in Draco and Ursa Minor are real MT-BSSs.

\section{Comparison with other galaxies and globular clusters}

BSSs have been observed in most globular clusters and at least in four
dSphs: Draco (A01), Sculptor (Hurley-Keller et
al. 1999), Sextans (L03), and Ursa Minor (C02). 
 It is instructive to compare our findings with previous papers on both
dSphs and globular clusters.

\subsection{Comparison with other dSphs}

Previous studies of Draco (A01) and Sculptor
(Hurley-Keller et al. 1999) do not report information about the radial
distribution or the luminosity of BSSs. However, Hurley-Keller et
al. (1999) calculate the ratio $N_{\rm BSS}/N_{\rm SGB}$ in the inner
region of the galaxy (a box of 15'$\times{}$15' centred on the centre of the galaxy) and in the outer
one. They find that this ratio changes only by a factor 1.5 between the
inner and the outer region, suggesting that BSSs are not very
concentrated.

On the other hand, A01 suggest that BSS candidates in Draco are
consistent with a population of intermediate age stars. In fact, they
find a 'red clump' population in the CMD diagram, which might support
this interpretation. This hypothesis cannot be ruled out also in the
case of our data (see discussion in Appendix~B).
% which hypothesis ? this is a bit unclear

C02 analysed the ratio of the number of `blue plume' stars (corresponding 
to a wider definition of BSS) and the number of HB stars as a function of 
radius (see their fig.~10) in Ursa Minor. They find an almost flat
distribution, whereas we note a small rise in the relative frequency of BSSs
around $1.4\,{}r_c$, but given the large error bars, our
distribution is also consistent with a flat one ($\chi{}^2\sim{}1$
%marginally ?
with 6 data points).  However, not only was the definition of BSS in C02
different, but also the observed photometric bands: C02 build their CMD
by plotting a {\it `V'} magnitude, which is actually the average
between $R$ and $B$ magnitudes, versus the ($B-R$) colour. Thus, our results
and those of C02 are not directly comparable. The most important fact is that
both C02 and our findings suggest that there is no central peak of BSSs
in Ursa Minor.

As we already mentioned in Section 3, L03 show both the radial and the
luminosity distribution of BSS candidates in Sextans. Furthermore, their
data are more easily comparable with ours, as they use the same
filters. However, even if we adopt the same definition of BSS and the
same normalization as L03, we do not find in either Draco or Ursa Minor
any correlation between the brightness and the radial position of the BSSs,
unlike that reported by L03 in Sextans. 
This discrepancy is unlikely due to the lack of statistics in our data, because, when we adopt the same selection criteria as L03, the number of BSSs rises to 198 in Draco and 212 in Ursa Minor, which is quite close to the sample of L03 (i.e. $\sim{}~230$ BSSs).
%(Fig.~\ref{fig:figlee}).

%In Fig.~\ref{fig:figboh} we report the BSS radial distribution in Draco (left) and Ursa Minor (right panel), adopting the L03
%definition and normalization  (but the results do not change if we adopt
%our definitions). 
%We do not observe any significant
%differences between faint ($22.1<V<22.8$) and bright ($20.9<V<22.1$)
%BSS. 
%Indeed, bright BSS in Draco seem slightly less concentrated than
%the faint ones.  The discrepancy between our results and that of L03
%could be explained  with an intrinsic difference of BSS in
%Sextans with respect to Draco and Ursa Minor (see Section 5).

%The distribution of $N_{{\rm BSS}}/N_{{\rm SGB}}$
%%($N_{{\rm SGB}}$ is the number of SGB stars per each annulus) 
%is very
%similar to the distribution of both $N_{{\rm BSS}}/N_{{\rm RGB}}$ and
%$N_{{\rm BSS}}/N_{{\rm HB}}$. These properties suggest 

The difference here might be due to statistical fluctuations, or could
simply be connected with the intrinsic properties of Sextans, which are quite
different from those of Draco and Ursa Minor. For example, Sextans has a
higher concentration index ($c\sim{}1$) with respect to both Draco and Ursa
Minor, and it is very extended ($r_c$= 16'.6 and $r_t=160$', IH95).  L03
show the radial distribution of BSSs only within $\sim{}1.1\,{}r_c$,
without information about external BSSs.  It has recently
been claimed that the centre of Sextans contains a
kinematically distinct stellar population, which might be associated
with a star cluster (Kleyna et al. 2004; but also see  Walker et al. 2006). 
Indeed, the correlation between position and brightness of BSSs could be 
explained by invoking the presence of a star cluster. In this case, the bright 
BSSs, more concentrated toward the centre, could be COL-BSSs or even young 
stars formed in
the star cluster; whereas the faint BSSs are MT-BSSs, like those in Draco
and Ursa Minor. (We note that the distribution of faint BSSs alone in
Sextans is quite similar to the distribution of the entire BSS sample in
Draco and Ursa Minor.)

\subsection{Comparison with globular clusters}
%%%%%%%%%%%%%%%%%%%%%%%%%%%%%%%%%%% FIGURE 8 %%%%%%%%%%%%%%%%%%%%%%%%%%%%%%%%%%
\begin{figure}
\center{{
\epsfig{figure=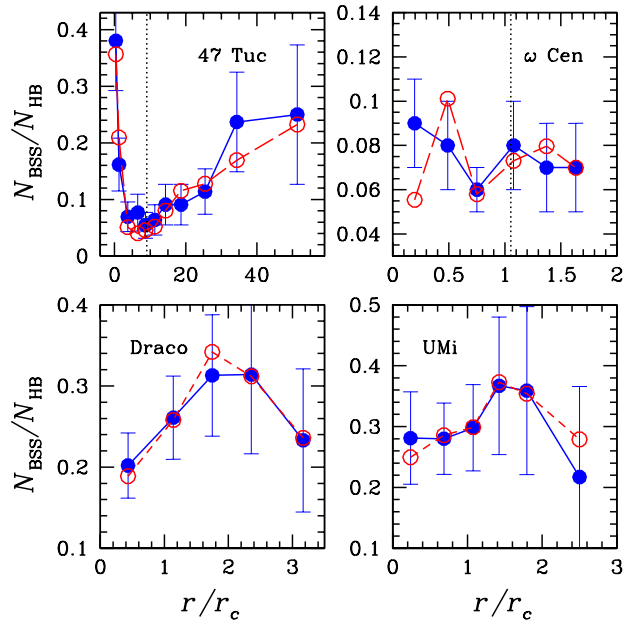,height=8.5cm}
}}
\caption{\label{fig:fig9}
From left to right and top to bottom: relative frequency of BSSs normalized to HB stars in 47~Tucanae, $\omega{}$ Centauri, Draco and Ursa Minor. Filled (open) circles connected by the solid (dashed) line are the observations (best-fitting simulations). The dotted vertical line in 47~Tucanae and  $\omega{}$ Centauri panels indicate $r_{av}$. The data and models for Draco and Ursa Minor are the same as in Fig.~\ref{fig:fig7} and Fig.~\ref{fig:fig8}, respectively. The data for 47~Tucanae and $\omega{}$ Centauri are from Ferraro et al. (2004) and from Ferraro et al. (2006), respectively. Models for 47~Tucanae and $\omega{}$ Centauri are from M06.
}
\end{figure}
%%%%%%%%%%%%%%%%%%%%%%%%%%%%%%%%%%%%%%%%%%%%%%%%%%%%%%%%%%%%%%%%%%%%%%%%%%%%%%%

What are the main differences between BSSs in dSphs and BSSs in
globular clusters? Are there any globular clusters whose BSSs behave like
those in dSphs? Fig.~\ref{fig:fig9} shows the distribution of $N_{\rm
BSS}/N_{\rm HB}$ of Draco and Ursa Minor together with that of
47~Tucanae and $\omega{}$ Centauri (from M06). We chose 47~Tucanae and
$\omega{}$ Centauri, because they have very different BSS populations.

47~Tucanae is a sort of prototype for BSSs in globular clusters: it
clearly shows the bimodal BSS relative frequency, which has been
observed in more and more globular clusters in the last few years
(Ferraro et al. 1993, 1997; Zaggia et al. 1997; Ferraro et al. 2004;
Sabbi et al. 2004; Warren, Sandquist \& Bolte 2006; Lanzoni et
al. 2007a). The clusters which do not have a bimodal distribution,
in general show only the central peak of BSSs, which rapidly drops outside
the core (see e.g. NGC1904; Lanzoni et al. 2007b).

The bimodal distribution might be explained by requiring that the
central peak is populated mainly by COL-BSSs, the  external
increase is due to MT-BSSs which have not yet sank to the centre, and
the minimum of the BSS distribution is connected with the efficiency of
dynamical friction (M04, M06). In fact, the position of the minimum has
been found to be equal to the maximum distance ($r_{av}$) from the
centre at which dynamical friction is able to bring binaries
(progenitors of MT-BSSs) into the core within the lifetime of the cluster
(M04, M06).  In this scenario, globular clusters without the external
rise (e.g. NGC1904) are expected to be poor in mass-transferring binaries,
or not to have formed MT-BSSs in the last Gyrs.

From Fig.~\ref{fig:fig9} it is clear that the distribution of BSSs in
Draco and Ursa Minor is completely different from that of a typical
globular cluster like 47~Tucanae. In particular, it seems that Draco and
Ursa Minor have only the peripheral rise of BSSs, and completely lack the
central peak. As we already discussed in Section~3, this supports the
idea that the central peak in globular clusters is due to COL-BSSs,
whereas the external rise is due to MT-BSSs.

Instead, $\omega{}$ Centauri is unique among the globular
clusters where BSSs have been already observed. In fact, $N_{\rm
BSS}/N_{\rm HB}$ and $N_{\rm BSS}/N_{\rm RGB}$ in $\omega{}$ Centauri
are both consistent with a flat distribution. M06 suggested that this
distribution might be the product of both the lack of COL-BSSs in the
core of $\omega{}$ Centauri (its core density being quite low)
and the inefficiency of dynamical friction. In fact, due to the joint
effect of a low central density ($\sim{}6\times{}10^3$ stars pc$^{-3}$)
and of a high velocity dispersion ($\sim{}17$ km s$^{-1}$), the
dynamical friction time-scale in $\omega{}$ Centauri is a factor of
$\sim{}$200 longer than in 47~Tucanae. As dynamical
friction is inefficient, binaries do not sink into the centre, and the
minimum  in the BSS distribution does not appear.

In this sense, $\omega{}$ Centauri appears as something midway between the other globular clusters and the
dSphs.  Different from dSphs, it can still form COL-BSSs in the
centre, because its core density is considerably higher than that of
dSphs; but its dynamical friction is inefficient in 
   moulding the shape of BSS distribution, exactly as in
dSphs. Fig.~\ref{fig:fig9} even suggests the idea of continuity between
47~Tucanae, $\omega{}$ Centauri and the two dSphs: as the central
density of the system decreases, the central peak disappears, and the
BSS distribution becomes less and less concentrated.

In line with this idea is the distribution of $r_{av}$ (see Fig.~\ref{fig:fig9}), which
in 47~Tucanae and in many other globular clusters is $\sim{}10\,{}r_c$,
in $\omega{}$ Centauri is $\sim{}1\,{}r_c$, while in Draco and Ursa
Minor it does not even appear in the plot, because it is consistent with
0 ($r_{av}\lesssim{}5\times{}10^{-2}\,{}r_c$).

 We note that, apart from the BSS distribution,
$\omega{}$ Centauri displays several features which are indicative of an object
midway between globular clusters and dSphs: the metallicity spread, the evidences for rotation, the large mass and the low concentration are quite atypical for a globular cluster; so that  some authors (Zinnecker et al. 1988;
Freeman 1993; Ideta \& Makino 2004) claim that $\omega{}$ Centauri is
not a real globular cluster, but the nuclear remnant of a dwarf galaxy.

%It is not a case if some  authors (Zinnecker et al. 1988;
%Freeman 1993; Ideta \& Makino 2004) claim that $\omega{}$ Centauri is
%not a real globular cluster, but the remnant of a dwarf galaxy.

Finally, in Fig.~\ref{fig:fig9} it is also apparent that the level of $N_{\rm BSS}/N_{\rm HB}$ ($\sim{}0.05-0.4$) in Draco and Ursa Minor is comparable with the level in 47~Tucanae and $\omega{}$ Centauri, analogous to most of the globular clusters (see e.g. M06). Thus, we can conclude that the fraction of BSSs versus HB stars in these two dSphs and in globular clusters are similar. This fact indirectly supports the hypothesis that BSS candidates in Ursa Minor and Draco are real MT-BSSs. In fact, if all the BSSs are  MT-BSSs,  $N_{\rm BSS}/N_{\rm HB}$  should
reflect both the ratio of lifetimes and the fraction of stars in suitable
binaries,
and should be constant for the same turn-off mass and metallicity populations, if
the binary fraction is constant.
%Interestingly, $\omega{}$ Centauri is different from the other clusters, with an uncommonly low  $N_{\rm BSS}/N_{\rm HB}$  (Ferraro et al. 2006).

\section{Summary}

In this paper we addressed the problem of BSS candidates in
dSphs in general, and in Draco and Ursa Minor in particular. There are
two fundamental open questions about BSSs in  dSphs: i) whether
they are authentic BSSs or young stars; and ii) what is their formation
mechanism?

We analysed both the radial and the luminosity distributions of these
stars, and we compared the data with dynamical simulations of BSSs.  The
main feature of the observed radial distribution of BSSs, normalized to
RGB or HB stars, is the absence of a central peak. Even if the young
% ??
stars' interpretation cannot be dismissed (at least for Draco; see
Appendix~B), this suggests that BSS candidates in Draco and Ursa Minor
are actually true BSSs.  Furthermore, the almost flat radial distribution
is consistent with theoretical models (M06) for MT-BSSs, i.e. BSSs which
formed by mass transfer in isolated binaries.  Also the luminosity
distribution, which does not show any correlation with the position of
BSSs, agrees with theoretical models of mass-transfer BSS formation.
%Finally, we found that the observed radial distribution agrees very well
%with dynamical simulations of MT-BSS.

These findings support the model by M04 and M06, which
explains the formation of BSSs by the joint contribution of stellar
collisions and mass transfer in isolated binaries. This model was 
originally developed only for globular
clusters, but we find that it works also for dSphs. As predicted by M06,
the presence of a central peak in the relative frequency of BSSs is due
to COL-BSSs, and can be explained only if both stellar collisions and
dynamical friction are efficient. The peak tends to disappear if the
central density of the system is too low and/or its dynamical friction
time is too long. This idea  was confirmed by the absence of any
central peak in $\omega{}$ Centauri, and  now we find that this
result is even stronger in Draco and Ursa Minor.

Low-density systems, where stellar collisions do not occur, can form
only MT-BSSs, whose initial distribution mirrors the distribution of the
progenitor binaries. The less efficient the dynamical friction, the more
the BSS distribution is similar to the distribution of progenitor
binaries.  This idea is
fully supported by Draco and Ursa Minor BSSs: the best-fitting
simulations are based on an isothermal distribution between
(approximately) the core and the tidal radius, as we would expect for a
distribution of primordial binaries.

 Furthermore, Momany et al. (2007) recently analysed the BSS candidates of 8 dSphs (Draco and Ursa Minor among them) and found a statistically significant anti-correlation between the relative frequency of BSS candidates ($N_{\rm BSS}/N_{\rm HB}$, calculated over the entire galaxy) and the total luminosity of the dSph. If BSS candidates were young MS stars rather than real BSSs, such anti-correlation would not make sense.

Thus, from our analysis as well as from  Momany et al. (2007) we conclude that BSS candidates in Draco and Ursa Minor  behave like real MT-BSSs, rather than young MS stars. This suggests (even if it does not definitely prove) that Draco and Ursa Minor are  'fossil' galaxies, where star formation was completely
suppressed many Gyrs ago.  This scenario is also confirmed by recent
simulations (Mayer et al. 2007), which indicate that Draco and Ursa
Minor, two of the closest dSphs to the Milky Way,
had all their gas removed  $\sim{}10$~Gyr ago, probably by tidal shocks and
ram pressure
exerted
by the Milky Way.  The 'fossil' nature of Draco and Ursa Minor would make
them a natural place to study the conditions at the earliest epochs of galaxy formation.

On the other hand, it would be interesting to throughly study the
nature of BSS candidates in other dSphs, like Sextans,
where star formation probably lasted longer. The main goal would be
to understand whether, and what fraction of, these stars are authentic
BSSs, in order to disentangle the history of BSS formation from that of
MS stars.

\section {Acknowledgments}
We thank the referee, T. Maccarone, for the critical reading of the
manuscript. We also thank F. D'Antona, F. Ferraro, S. Zaggia and
Y. Momany for useful discussions. MM acknowledges support from the Swiss
National Science Foundation, project number 200020-109581/1
(Computational Cosmology \&{} Astrophysics). ER acknowledges support from the
Netherlands Organization for Scientific Research (NWO) under project
number 436016. MM and ER thank the Kapteyn Astronomical Institute of the
University of Groningen, and
the Institute for Theoretical Physics of the University of Z\"urich for
the hospitality during the preparation of this paper.

\appendix

\section{Foreground subtraction}

%%%%%%%%%%%%%%%%%%%%%%%%%%%%%%%%%%% FIGURE A1 %%%%%%%%%%%%%%%%%%%%%%%%%%%%%%%%%
\begin{figure}
\center{{
\epsfig{figure=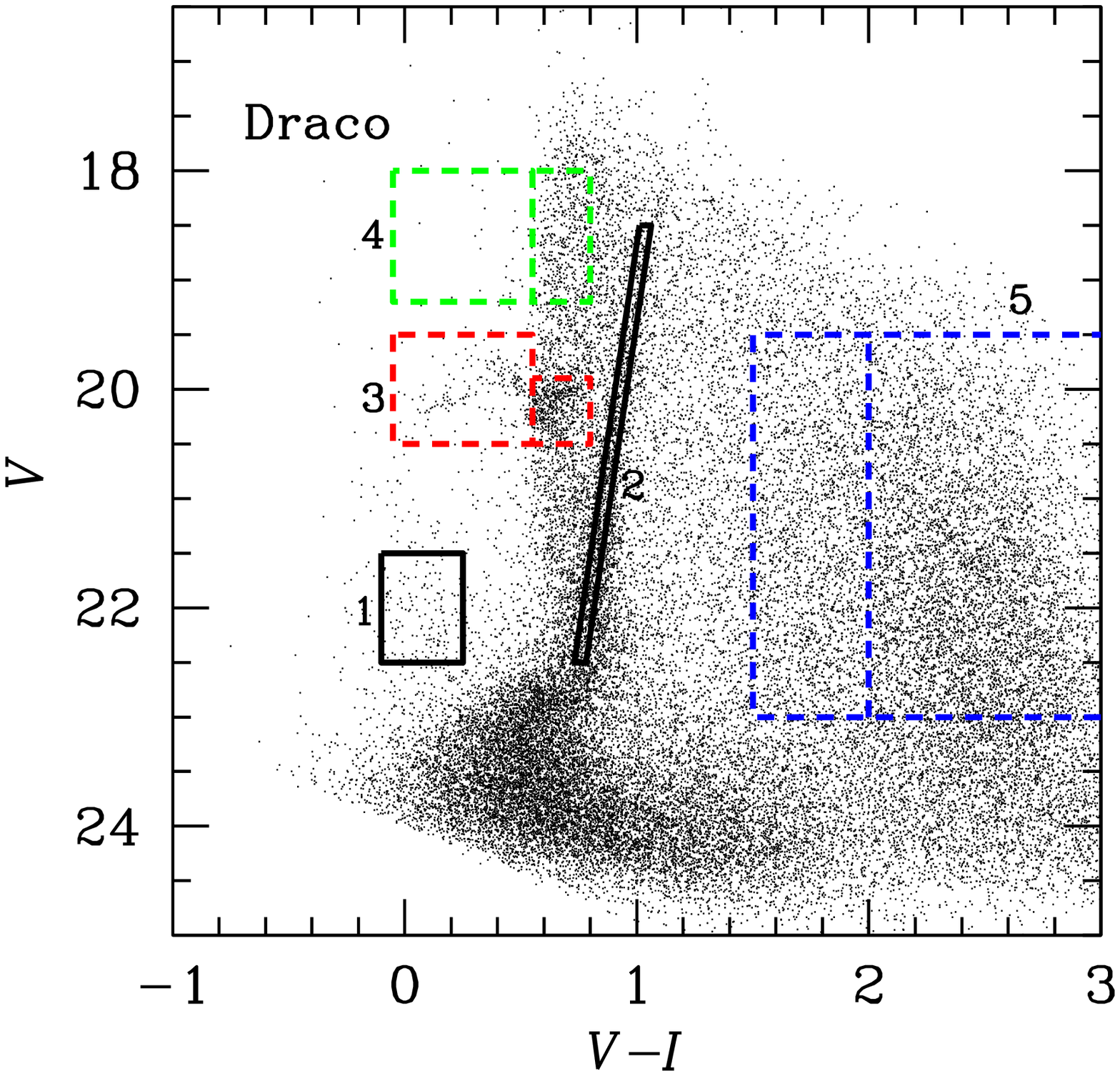,height=8.5cm}
}}
\caption{\label{fig:cmd_draco_full} 
CMD diagram of Draco. As in Fig. \ref{fig:fig1}, the
selection areas for BSS (1), RGB (2), and HB (3) are shown. We also
show the selection boxes for the HB foreground (4) and the VRS (5).}
\end{figure}
%%%%%%%%%%%%%%%%%%%%%%%%%%%%%%%%%%%%%%%%%%%%%%%%%%%%%%%%%%%%%%%%%%%%%%%%%%%%%%%

The contamination of the data due to foreground Milky Way stars  (and
also to background objects) is quite evident, especially for the RGB
region (see Figs. \ref{fig:fig1} and \ref{fig:cmd_draco_full}).
%\footnote{{\bf [ER] The CMD diagrams shown in
%Fig. \ref{fig:fig1} appear to have a different level of contamination
%only because of the spatial extent of the available data: for Draco, the
%observations covered an area of the sky which extends well beyond the
%tidal radius; for Ursa Minor, observations are available only within
%about one tidal radius}}
Removing such contamination is
important, especially in the outer regions of the dSphs.

We note that, although we refer only to `foreground'
removal, the methods outlined below work equally well for the
subtraction of the contamination by compact background objects.

%, even if
%this operation is not mentioned in previous papers (e.g. A01).

%In order to avoid foreground contamination, first of all we consider foreground all those stars which are outside $r_{ext}=45$' () from the centre of Draco (Ursa Minor). These stars are well outside the tidal radius, $r_t$, whose value can be derived from Table~1. Then, we subtract the foreground component also for stars within  $r_{ext}$ by adopting the following method.

\subsection{RGB and BSS foreground contamination}

In order to estimate the foreground contamination for RGB stars and BSSs, we
adopt the following method, both in Draco and in Ursa Minor.

First of all, we assume that all stars redder than ($V-I$)=1.5 and with $V$
in the [19.5, 23] range (hereafter VRS, i.e. very red stars) are in the
foreground. In fact, the VRS region of the CMD (box 5 in
Fig. \ref{fig:cmd_draco_full}) should not be populated by stars
belonging to Draco, nor to Ursa Minor\footnote{We also checked a
more restrictive definition of VRS, i.e. ($V-I$)$\geq$2.0 and the same $V$
magnitude range. No significant difference was found in our results.}.

Second, we expect that all the stars which are outside $r_t$ do not
belong to the dSph, independent of their colour and magnitude. This is
useful because the Draco data extend well beyond $r_t$, and we can
select a subset of `external' stars, which we define to include all
the stars whose elliptical radial distance from the centre exceeds
$r_{t}$=45.1'.
%\footnote{Ideally, $r_{ext}\simeq r_t$=28.3'. But in the
%elliptical corona between 28.3' and 45' there is still some
%contamination from Draco stars; this was already reported in previous
%studies, even if it is not clear whether we are in presence of
%extra-tidal stars (IH95), or the tidal radius was underestimated
%(e.g. A01 estimate a tidal radius of 42' for Draco).}.

We can therefore count the number of VRS, RGB and BSS equivalents\footnote{Here we use the
name of BSSs and RGB just for convenience. These are foreground
stars which happen to have the same colour and magnitude of BSSs and RGB,
respectively.} with $r_{ell}>r_{t}$ ($N_{{\rm VRS},\,{}ext}$, $N_{{\rm
RGB},\,{}ext}$, and $N_{{\rm BSS},\,{}ext}$, respectively), and we
derive the ratios
$f_{{\rm RGB}/{\rm VRS}} = N_{{\rm RGB},\,{}ext}/N_{{\rm VRS},\,{}ext}
\simeq 0.0318\pm0.0025$,
and
$f_{{\rm BSS}/{\rm VRS}} = N_{{\rm BSS},\,{}ext}/N_{{\rm VRS},\,{}ext}
\simeq 0.0031\pm0.0008$.

These ratios should be independent of position, as we have tested this in two ways.
First of all, we looked for fluctuations of the surface density of VRS
stars as a function of radius. Although small fluctuations are present,
the overall density can be considered constant in the whole Draco field (it is
consistent with a constant value of 0.94 VRS/arcmin$^2$, with reduced
$\chi{}^2\simeq$0.8 over 23 radial bins).
% You could argue that 2 is significantly different from 1 (~4 sigma). 
% Extinction varies noticeably over the Draco region have you corrected for
% this doing the tests ?  I have an extinction-corrected version of the
% of the catalogue if required.
As a further test, we split the `external' region into its eastern
and western half and checked that there is no statistically significant
difference between the values of $f_{{\rm RGB}/{\rm VRS}}$ and of
$f_{{\rm BSS}/{\rm VRS}}$ which were obtained in the two halves. 

The foreground contamination of the $i$-th elliptical annulus can be
estimated by counting the number $N_{{\rm VRS},i}$ of VRS stars in the
annulus, and converting it into the expected number of foreground
BSSs (RGB stars) through the factor
$f_{{\rm BSS}/{\rm VRS}}$ ($f_{{\rm RGB}/{\rm VRS}}$). Then, the corrected
number of BSSs (RGB stars) is simply
\begin{eqnarray}
N_{{\rm BSS},i} & = \; N_{{\rm BSS,obs},i} & - \quad
N_{{\rm VRS},i}\,{} f_{{\rm BSS}/{\rm VRS}}\\
N_{{\rm RGB},i} & = \; N_{{\rm RGB,obs},i} & - \quad
N_{{\rm VRS},i}\,{} f_{{\rm RGB}/{\rm VRS}},
\end{eqnarray}
where $N_{{\rm BSS,obs},i}$ ($N_{{\rm RGB,obs},i}$) is the number of
BSSs (RGB stars) observed in the annulus.

In the case of Ursa Minor, we do not have enough data at large
radial distances from the centre, and therefore cannot obtain a local estimate
for $f_{{\rm RGB}/{\rm VRS}}$ and $f_{{\rm BSS}/{\rm VRS}}$. For this
reason, we use the values obtained for Draco also for Ursa Minor. This
is not optimal, but not unreasonable, as the two dSphs are at comparable
Galactic latitudes.

\subsection{HB foreground contamination}

The above procedure for RGB and BSSs could also be used  for HB stars.
However, for HB stars we adopt a more straightforward technique, making use
of the fact that the foreground does depend on colour, whereas it is
nearly independent of magnitude (at least in the range considered in our CMD).

Such a fact cannot be exploited in the case of RGB and BSSs, because it
requires a CMD region which is both in the same colour range as BSSs or
RGB, and is populated by foreground stars only.  However, for HB stars, the
dashed regions labelled by 4 in Fig.~\ref{fig:cmd_draco_full} are at exactly the
same ($V-I$) range of the regions (labelled 3) where HB stars are selected, and
are almost exclusively populated by foreground stars.

In order to account for the different foreground level for RHB and  BHB stars,
we further divided region 4 of Fig.~\ref{fig:cmd_draco_full} into two sub-regions:
a blue one with the same colour range of BHB, and a red one with the same
colours of RHB. We will refer to stars in the two sub-regions as to the
fgBHB, and the fgRHB stars, respectively.

Then, the corrected numbers of RHB (BHB) stars in the $i$-th annulus are
\begin{eqnarray}
N_{{\rm RHB},i} & = \; N_{{\rm RHB,oss},i} -
N_{{\rm fgRHB},i} & {A_{\rm RHB}\over A_{\rm fgRHB}}\\
N_{{\rm BHB},i} & = \; N_{{\rm BHB,oss},i} -
N_{{\rm fgBHB},i} & {A_{\rm BHB}\over A_{\rm fgBHB}},
\end{eqnarray}
where $N_{{\rm RHB,oss},i}$($N_{{\rm BHB,oss},i}$) is the number of
RHB (BHB) stars which was actually observed in the annulus,
$N_{{\rm fgRHB},i}$($N_{{\rm fgBHB},i}$) is the number of fgRHB (fgBHB) stars
in the annulus, and $A_{\rm RHB}/A_{\rm fgRHB}$
($A_{\rm BHB}/A_{\rm fgBHB}$) is a correction factor which accounts for
the different extensions of the various regions in the CMD.

Foreground subtraction is then carried out by subtracting the number of
fgBHB (fgRHB) stars in the annulus (after a correction accounting for the
ratios of the CMD areas) from the number of BHB (RHB) stars in the annulus. 

We note that this method of foreground subtraction has a slight
dependence on the radial distance, the foreground level within the core
radius being generally higher than outside. This is because, in addition
to subtracting Milky Way stars in the foreground component, this technique also
accounts for extra-effects (like binaries, blending, errors in the
observed magnitude and contamination from other stellar types, like
RGB).

As a sanity check, we compared the distribution of HB stars obtained by using
this method of foreground subtraction and that obtained using the same
procedure as for RGB and BSSs. The differences in the inner annuli are
negligible. In the outer annuli ($>2\,{}r_c$, where $r_c$ is the core
radius) the difference is larger, but remains within the (quite large)
Poissonian error bars.

%%%%%%%%%%%%%%%%%%%%%%%%%%%%%%%%%%%%%%%%%%%%%%%%%

\section{A test of the young star hypothesis through isochrones}

%%%%%%%%%%%%%%%%%%%%%%%%%%%%%%%%%%% FIGURE B1 %%%%%%%%%%%%%%%%%%%%%%%%%%%%%%%%%%
\begin{figure*}
\center{{
\epsfig{figure=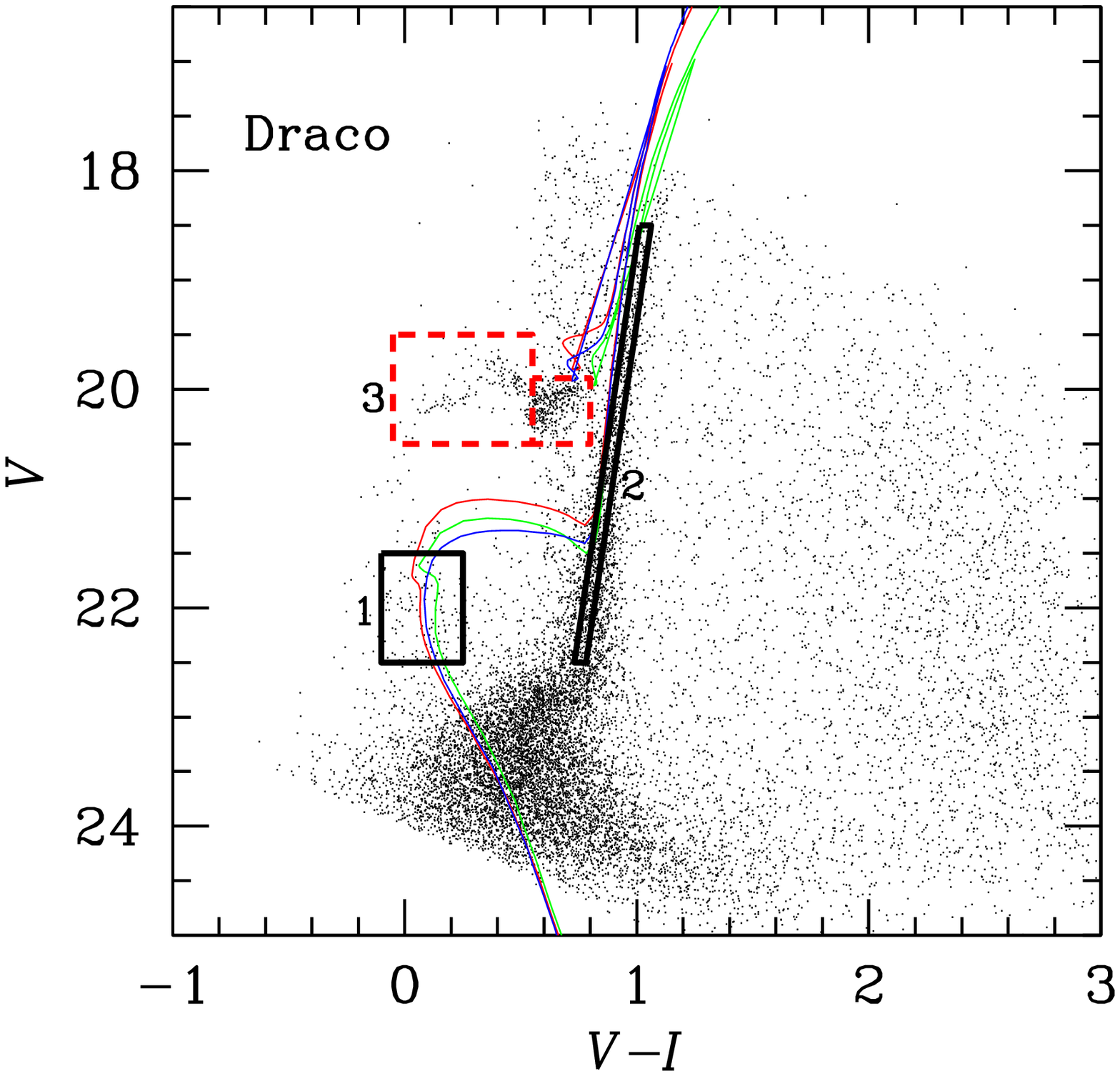,height=8cm}
\epsfig{figure=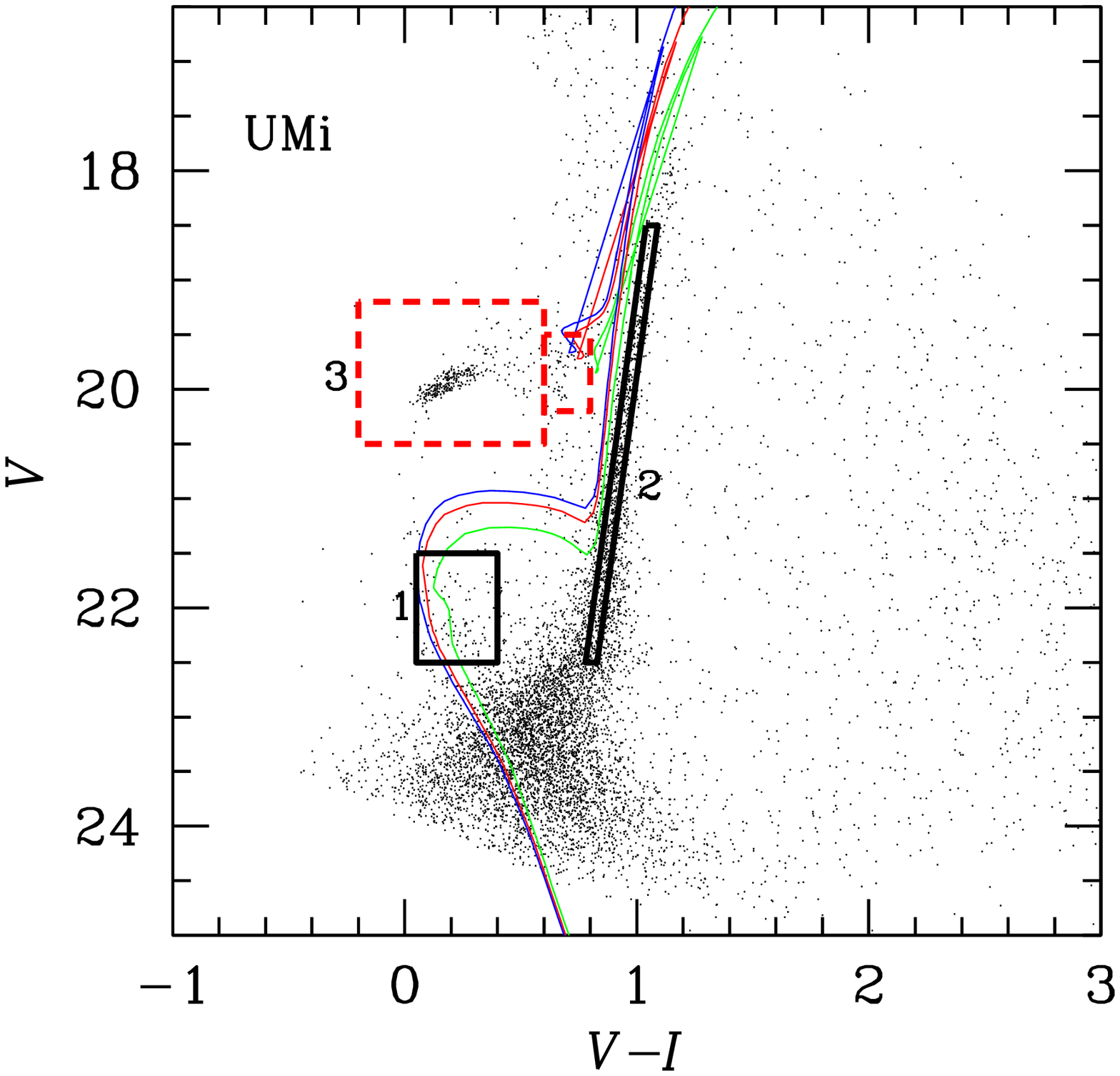,height=8cm}
}}
\caption{\label{fig:figb1} Reddening and distance corrected isochrones
of single stellar populations superimposed to the CMD of the central
region of Draco (left panel; here, in order to reduce foreground
contamination, we plot only stars with $r\leq28.3'$) and Ursa Minor
(right). In both cases, the left line refers to a metallicity
[Fe/H]=-2.3, the middle one to [Fe/H]=-2.0, and the right one to
[Fe/H]=-1.5. We assumed ages of 2.0 and 2.5 Gyr in the case of Draco and
Ursa Minor, respectively. Boxes are the same as in Fig.~\ref{fig:fig1}.}
\end{figure*}
%%%%%%%%%%%%%%%%%%%%%%%%%%%%%%%%%%%%%%%%%%%%%%%%%%%%%%%%%%%%%%%%%%%%%%%%%%%%%%%

Although we have shown that the properties of the observed BSS
population are fully compatible with the expectations for `real' BSSs,
the `young star' interpretation provided by A01 for Draco still
remains viable.

A more direct test of the nature of this population can be performed by
looking for other hints of a relatively young population. Actually, the
A01 interpretation of the BSSs in Draco was based on the
observation of a small concentration of stars in a region of their CMD
(the `red clump', i.e. region 13 of their fig. 13) which should not
be populated if no star formation occurred in the last 10 Gyr.

We performed a similar test by means of the isochrones of the Padova
group (see Girardi et al. 2002 ; see also
{\tt http://stev.oapd.inaf.it/$\sim{}$lgirardi/cmd}).

We plotted a set of theoretical isochrones over the Draco and Ursa Minor
CMDs\footnote{We assumed a reddening E($B-V$)=0.03 for both galaxies, a
value for which a vast consensus exists. The distance modulus of Ursa
Minor was chosen to be 19.41 [Bellazzini et al. (2002) and C02 found
19.41$\pm$0.12 and 19.40$\pm$0.10, respectively]. The distance
modulus of Draco is more controversial, as recent determinations yielded
relatively different values: A01, and Bonanos et al. (2004) found
compatible values (19.5$\pm$0.2 and 19.40$\pm$0.15$\pm$0.02,
respectively); but Bellazzini et al. (2002) found a significantly higher
value (19.84$\pm$0.14). We adopted the intermediate value of 19.60. We cannot make an estimate from our data, as the tip of the RGB in our WFC data is beyond the saturation limit.},
% is you include the -9 category by the way you will populate the upper part
% of the CMDs.  In forming the band-merge cataloge a correction is made for
% saturation which extends the dynamic range by about 1-2 mags.
varying both the age and the metallicity of the stellar population (see
Fig. B1 for some examples). From such isochrones it is clear that BSSs
lie close to the isochrones describing low-metallicity
($[Fe/H]\leq-1.5$, perfectly compatible with current estimates for Draco
and Ursa Minor) stellar populations with ages between 2 and 3 Gyr.

For Draco we chose to combine a Chabrier (2003) log-normal IMF with an
isochrone for an age of 2 Gyr and a metallicity [Fe/H]=-2.0, in order to
estimate the number of stars which should be expected in other regions
of the CMD if all of the observed BSSs are actually part of an
intermediate-age population. Within this scenario, the regions where we
expect the maximum number of intermediate-age stars and the minimum
contamination from the old population, are:\begin{enumerate}
\item{}{the young MS just below the BSS selection box;}
\item{}{the bright and faint part of of the BSS selection box;}
\item{}{the red clump.}
\end{enumerate}
In Table~B1 we list the theoretical predictions from the
isochrones ($N_{\rm pred}$), the total number of stars observed in each
CMD region ($N_{\rm raw}$), the estimated foreground contamination
($N_{\rm fg}$), and the number of observed stars after the foreground
subtraction ($N_{\rm obs}$).

%% It is clear that the predictions from the young star hypothesis are
%% quite compatible with our observations of Draco: the ratio of faint BSSs
%% to bright BSSs is slightly lower than expected, but the difference is
%% just at the $1\sigma$ level; on the other hand, the predicted number of
%% young MS and red clump stars is perfectly compatible with observations,
%% as both these CMD regions are likely to be contaminated by the old Draco
%% stellar population (old MS stars close to the turn-off for the young MS
%% region, HB and RGB stars for the red clump region).

It is clear that the predictions from the young star hypothesis are
quite compatible with our observations of Draco: the ratio of faint BSSs
to bright BSSs is slightly lower than expected ($1.75\pm0.49$ instead of
$2.29$), but the difference is just at the $1-\sigma$ level; on the
other hand, the predicted number of young MS and red clump stars is
perfectly compatible with observations, as both these CMD regions are
likely to be contaminated by the old Draco stellar population (old MS
stars close to the turn-off for the young MS region, RGB and especially
- given the partial superposition of the two regions - HB stars for the
red clump region).

%% We applied the same method also to Ursa Minor, using an isochrone age
%% of 2.5 Gyr. Results are summarized in Table B2, where we omitted the
%% Young MS region because of the very strong contamination from the old
%% MS. The main result is that the number of stars in the red clump region
%% is significantly lower than expected from the young stars hypothesis:
%% the predicted 9.4 red clump stars are within the $1\sigma$ upper limit
%% from the observations (10.4), but contamination from HB and RGB stars is
%% surely present, and there remains very little space for the young star
%% hypothesis. However, we note that a distance modulus 0.1-0.2
%% magnitude larger than what we assumed could ``hide'' the red clump stars
%% within the HB.

We applied the same method also to Ursa Minor, using an isochrone age of
2.5 Gyr. Results are summarized in Table B2, where we omitted the Young
MS region because of the very strong contamination from the old MS. The
number of stars in the red clump region appears to be extremely close to
the prediction from the young stars hypothesis, but a strong
contamination is surely present, as the red clump selection box
($19.62\ge V\ge 19.12$, $0.90\ge (V-I) \ge 0.50$) largely superimposes
with the HB (see Fig. B1). Such a strong contamination accounts for most
of the `excess' stars in the considered CMD region. However, the young
star hypothesis might still be viable, because most of the red clump
stars might be `hidden' within the HB.

In summary, the interpretation that BSS candidates in Draco are
intermediate-age stars can neither be ruled out, nor be confirmed by the
isochrone method applied to our observations. Only a spectral analysis
of stars in the red clump region could solve the uncertainty. In Ursa
Minor such an interpretation is hardly compatible with current data (see
also C02), but cannot be completely ruled out.

We point out that in both galaxies the mass of the intermediate-age
population needed to explain the BSS candidates is just about
$10^4\Msun$, which is a very small fraction of the mass of Draco or
Ursa Minor. If the age spread is $\gtrsim$ 1 Gyr, the implied star
formation rate is $\lesssim{}10^{-5}\,\Msun\,{}{\rm yr}^{-1}$, comparable
to the estimates shown in figs. 14 and 16 of A01, and much lower
than any observed star formation rate in dwarf galaxies.

%%%%%%%%%%%%%%%%%%%%%%%%%%%%%%% TABLE B1%%%%%%%%%%%%%%%%%%%%%%%%%%%%%%%%%
\begin{table}
\label{isoc_table_draco}
\begin{center}
\caption{Comparison of isochrone predictions (age 2.0 Gyr, [Fe/H]=-2.0)
with observations for Draco} \leavevmode
\begin{tabular}[!h]{lcccc}
\hline
CMD region
& $N_{\rm pred}$
& $N_{\rm raw}$
& $N_{\rm fg}$
& $N_{\rm obs}$\\
\hline
Young MS   & 36.8$\pm$3.4  & 43  & 2.3  & 40.7$\pm$7\\
BSS faint  & 84.2$\pm$7.7 & 76  & 2.3  & 73.7$\pm$9\\
BSS bright & 36.8$\pm$3.4  & 45  & 3.0  & 42.0$\pm$7\\
Red clump  & 15.3$\pm$1.4  & 101 & 53.3 & 47.7$\pm$11\\
\noalign{\vspace{0.1cm}}
\hline
\end{tabular}
\end{center}
\end{table}
%%%%%%%%%%%%%%%%%%%%%%%%%%%%%%%%%%%%%%%%%%%%%%%%%%%%%%%%%%%%%%%%%%%%%%%%%%%%%
%%%%%%%%%%%%%%%%%%%%%%%%%%%%%%% TABLE B2%%%%%%%%%%%%%%%%%%%%%%%%%%%%%%%%%
\begin{table}
\label{isoc_table_ursa}
\begin{center}
\caption{Comparison of isochrone predictions (age 2.5 Gyr, [Fe/H]=-2.0)
with observations for Ursa Minor} \leavevmode
\begin{tabular}[!h]{lcccc}
\hline
CMD region
& $N_{\rm pred}$
& $N_{\rm raw}$
& $N_{\rm fg}$
& $N_{\rm obs}$\\
\hline
BSS faint  & 63.6$\pm$6.4 & 71 & 1.0  & 70.0$\pm$9\\
BSS bright & 36.3$\pm$3.7 & 29 & 1.0  & 28.0$\pm$6\\
Red clump  & 11.4$\pm$1.2 & 28 & 16.2 & 11.8$\pm$6\\
\noalign{\vspace{0.1cm}}
\hline
\end{tabular}
\end{center}
\end{table}
%%%%%%%%%%%%%%%%%%%%%%%%%%%%%%%%%%%%%%%%%%%%%%%%%%%%%%%%%%%%%%%%%%%%%%%%%%%%%

Finally, the isochrones can also be used to give an indicative estimate
of the upper/lower limit mass of BSSs, which are used to set up our
simulations (see Section 4). For Draco we find that their masses should
be in the range $1.11-1.35\Msun$, whereas for Ursa Minor this range
moves slightly to $1.09-1.34\Msun$.

\end{document}